\documentclass[prb,aps,twocolumn,home,floats]{revtex4}

\usepackage{graphics}
\usepackage{citesort}
\usepackage{dcolumn}

\begin{document}

\title{{\rm\small\hfill to appear in: Handbook of Materials Modeling,\quad\quad\,\,}\\  
{\rm\small\hfill Vol. {\bf 1} Fundamental Models and Methods, Sidney Yip (Ed.)}\\
\quad\\
\boldmath $Ab \; initio$ atomistic thermodynamics and statistical
mechanics\\ of surface properties and functions}

\author{Karsten Reuter$^{1}$}
\author{Catherine Stampfl$^{1,2}$}
\author{Matthias Scheffler$^{1}$}

\affiliation{$^1$ Fritz-Haber-Institut der Max-Planck-Gesellschaft,
             Faradayweg 4-6, D-14195 Berlin, Germany}

\affiliation{$^2$ School of Physics, The University of Sydney,
             Sydney 2006, Australia}

\begin{abstract}
Previous and present ``academic'' research aiming at atomic scale understanding is mainly concerned with the study of individual molecular processes possibly underlying materials science applications. In investigations of crystal growth one would for example study the diffusion of adsorbed atoms at surfaces, and in the field of heterogeneous catalysis it is the reaction paths of adsorbed species that is analyzed. Appealing properties of an individual process are then frequently discussed in terms of their direct importance for the envisioned material function, or reciprocally, the function of materials is somehow believed to be understandable by essentially one prominent elementary process only. What is often overlooked in this approach is that in macroscopic systems of technological relevance typically a large number of distinct atomic scale processes take place. Which of them are decisive for observable system properties and functions is then not only determined by the detailed individual properties of each process alone, but in many, if not most cases also the interplay of all processes, i.e. how they act together, plays a crucial role. For a {\em predictive materials science modeling with microscopic understanding}, a description that treats the statistical interplay of a large number of microscopically well-described elementary processes must therefore be applied. Modern electronic structure theory methods such as density-functional theory (DFT) have become a standard tool for the accurate description of the individual atomic and molecular processes. In what follows we discuss the present status of emerging methodologies which attempt to achieve a (hopefully seamless) match of DFT with concepts from statistical mechanics or thermodynamics, in order to also address the interplay of the various molecular processes. The new quality of, and the novel insights that can be gained by, such techniques is illustrated by how they allow the description of crystal surfaces in contact with realistic gas-phase environments, which is of critical importance for the manufacture and performance of advanced materials such as electronic, magnetic and optical devices, sensors, lubricants, catalysts and hard coatings.
\end{abstract}
\maketitle

\begin{figure}
\scalebox{0.35}{\includegraphics{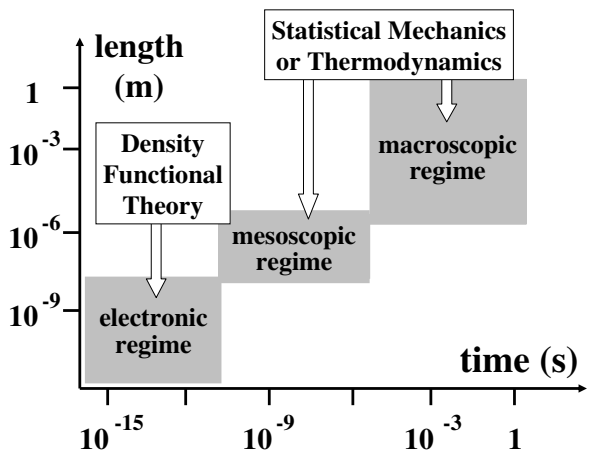}}
\caption{\label{fig_multiscale}
Schematic presentation of the time and length scales relevant for most material science applications. The elementary molecular processes, which rule the behavior of a system, take place in the so-called ``electronic regime''. Their interplay, which frequently determines the functionalities however, only develops after meso- and macroscopic lengths or times.}
\end{figure}

For obtaining an understanding, and for the design, advancement or refinement of modern technology that controls many (most) aspects of our life, a large range of time and length scales needs to be described, namely, from the electronic (or microscopic/atomistic) to the macroscopic, as illustrated in Figure \ref{fig_multiscale}. Obviously, this calls for a multi-scale modeling, were corresponding theories (i.e. from the electronic, mesoscopic, and macroscopic regimes) and their results need to be linked appropriately. For each length and time scale regime alone, a number of methodologies are well established. It is however, the appropriate linking of the methodologies that is only now evolving. Conceptually quite challenging in this hierarchy of scales are the transitions from what is often called a micro- to a mesoscopic-system description, and from a meso- to a macroscopic-system description. Due to the rapidly increasing number of particles and possible processes, the former transition is methodologically primarily characterized by the rapidly increasing importance of statistics, while in the latter, the atomic substructure is finally discarded in favor of a continuum modeling. In this contribution we will concentrate on the micro- to mesoscopic-system transition, and correspondingly discuss some possibilities of how atomistic electronic structure theory can be linked with concepts and techniques from statistical mechanics and thermodynamics. 

Since our aim is a materials science modeling that is based on understanding, predictive, and applicable to a wide range of realistic conditions (e.g. realistic environmental situations of varying temperatures and pressures), this mostly excludes the use of empirical or fitted parameters -- both at the electronic and at the mesoscopic level, as well as in the matching procedure itself. Electronic theories that do not rely on such parameters are often referred to as {\em first-principles} (or in latin: {\em ab initio}) techniques, and we will maintain this classification also for the linked electronic-statistical methods. Correspondingly, our discussion will mainly (nearly exclusively) focus on such {\em ab initio} studies, although mentioning some other work dealing with important (general) concepts. Furthermore, this chapter does not (or only briefly) discuss equations; instead the concepts are demonstrated (and illustrated) by selected, typical examples. Since many (possibly most) aspects of modern material science deal with surface or interface phenomena, the examples are from this area, addressing in particular surfaces of semiconductors, metals, and metal oxides. Apart from sketching the present status and achievements, we also find it important to mention the difficulties and problems (or open challenges) of the discussed approaches. This can however only be done in a qualitative and rough manner, since the problems lie mostly in the details, the explanations of which are not appropriate for such a chapter.

To understand the elementary processes ruling the materials science context, microscopic theories need to address the behavior of electrons and the resulting interactions between atoms and molecules (often expressed in the language of chemical bonds). Electrons move and adjust to perturbations on a time scale of femtoseconds (1 fs = $10^{-15}$\,s), atoms vibrate on a time scale of picoseconds (1 ps = $10^{-12}$\,s), and individual molecular processes take place on a length scale of 0.1 nanometer (1 nm = $10^{-9}$\,m). Because of the central importance of the electronic interactions, this time and length scale regime is also often called the ``electronic regime'', and we will use this term here in particular, in order to emphasize the subtle difference between {\em ab initio} electronic and semi-empirical microscopic theories. The former explicitly treat the electronic degrees of freedom, while the latter already coarse-grain over them and directly describe the atomic scale interactions by means of interatomic potentials. Many materials science applications depend sensitively on intricate details of bond breaking and making, which on the other hand are often not well (if at all) captured by existing semi-empiric classical potential schemes. A predictive first-principles modeling as outlined above must therefore be based on a proper description of molecular processes in the ``electronic regime'', which is much harder to accomplish than just a microscopic description employing more or less guessed potentials. In this respect we find it also appropriate to distinguish the electronic regime from the currently frequently cited ``nanophysics'' (or better ``nanometer-scale physics''). The latter deals with structures or objects of which at least one dimension is in the range 1-100\,nm, and which due to this confinement exhibit properties that are not simply scalable from the ones of larger systems. Although the molecular processes in the electronic regime operate on a sub-nanometer length scale, they do thus not necessarily fall into the ``nanophysics'' category (unless they take place in a nano-scale system and exhibit qualitatively new properties).

Although already quite involved, the detailed understanding of individual molecular processes arising from electronic theories is, however, often still not enough. As mentioned above, in many cases the system functionalities are determined by the concerted interplay of many elementary processes, not only by the detailed individual properties of each process alone. It can for example very well be that an individual process exhibits very appealing properties for a desired application, yet the process may still be irrelevant in practice, because it hardly ever occurs within the ``full concert'' of all possible molecular processes. Evaluating this ``concert'' of elementary processes one obviously has to go beyond separate studies of each microscopic process. Taking the interplay into account, however, naturally requires the treatment of larger system sizes, as well as an averaging over much longer time scales. The latter point is especially pronounced, since many elementary processes in material sciences are activated (i.e. an energy barrier must be overcome) and thus rare. This means that the time between consecutive events can be orders of magnitude longer than the actual event time itself. Instead of the above mentioned electronic time regime, it can therefore be necessary to follow the time evolution of the system up to seconds and longer in order to arrive at meaningful conclusions concerning the effect of the statistical interplay. Apart from the system size, there is thus possibly the need to bridge some twelve orders of magnitude in time which puts new demands on theories that are to operate in the corresponding mesoscopic regime. And also at this level, the {\em ab initio} approach is much more involved than an empirical one because it is not possible to simply ``lump together'' several not further specified processes into one effective parameter. Each individual elementary step must be treated separately, and then combined with all the others within an appropriate framework. 

\begin{table*}[t]
\vspace{.5cm}
\begin{tabular}{l @{\hspace*{1cm}} c @{\hspace*{1cm}} c @{\hspace*{1cm}} c}
\hline \hline
& Information & Time scale & Length scale \\
\hline 
Density-functional theory  & Microscopic & - & 
$\raisebox{.6ex}[-.6ex]{$<\!\!\!\!$}\raisebox{-.6ex}[.6ex]{$\sim$}\, 10^3$
atoms\\
{\em Ab initio} atomistic thermodynamics & Microscopic & Averaged & 
$\raisebox{.6ex}[-.6ex]{$<\!\!\!\!$}\raisebox{-.6ex}[.6ex]{$\sim$}\, 10^3$
atoms\\
{\em Ab initio} molecular dynamics & Microscopic & 
 $t\, \raisebox{.6ex}[-.6ex]{$<\!\!\!\!$}\raisebox{-.6ex}[.6ex]{$\sim$}\, 
$ 50 ps & $\raisebox{.6ex}[-.6ex]{$<\!\!\!\!$}\raisebox{-.6ex}[.6ex]{$\sim$}\, 10^3$ atoms\\
Semi-empirical molecular dynamics & Microscopic & 
$t \, \raisebox{.6ex}[-.6ex]{$<\!\!\!\!$}\raisebox{-.6ex}[.6ex]{$\sim$}\, $ 1 ns & $\raisebox{.6ex}[-.6ex]{$<\!\!\!\!$}\raisebox{-.6ex}[.6ex]{$\sim$}\, 10^3$ 
atoms\\
Kinetic Monte Carlo simulations & Micro- to mesoscopic & 1 ps \,
$\raisebox{.6ex}[-.6ex]{$<\!\!\!\!$}\raisebox{-.6ex}[.6ex]{$\sim$}\,  t \,
\raisebox{.6ex}[-.6ex]{$<\!\!\!\!$}\raisebox{-.6ex}[.6ex]{$\sim$}\, $ 1 hour & 
$\raisebox{.6ex}[-.6ex]{$<\!\!\!\!$}\raisebox{-.6ex}[.6ex]{$\sim$}\, $
1 $\mu$m \\ 
Rate equations & Averaged   & 0.1 s \,
$\raisebox{.6ex}[-.6ex]{$<\!\!\!\!$}\raisebox{-.6ex}[.6ex]{$\sim$}\, t \,
\raisebox{.6ex}[-.6ex]{$<\!\!\!\!$}\raisebox{-.6ex}[.6ex]{$\sim$}\,
 \infty$ & $\raisebox{.6ex}[-.6ex]{$>\!\!\!\!$}\raisebox{-.6ex}[.6ex]{$\sim$}\,$ 10 nm \\ 
Continuum equations & Macroscopic & 1 s \,
$\raisebox{.6ex}[-.6ex]{$<\!\!\!\!$}\raisebox{-.6ex}[.6ex]{$\sim$}\, t \,
 \raisebox{.6ex}[-.6ex]{$<\!\!\!\!$}\raisebox{-.6ex}[.6ex]{$\sim$}\, \infty$ & 
$\raisebox{.6ex}[-.6ex]{$>\!\!\!\!$}\raisebox{-.6ex}[.6ex]{$\sim$}\,$ 10 nm\\ 
\hline
\hline
\end{tabular}
\vspace{.5cm}
\caption{The time and length scales typically handled by different theoretical approaches to study chemical reactions and crystal growth.}
\label{approaches}
\end{table*} 

Methodologically, the physics in the electronic regime is best described by electronic structure theories, among which density-functional theory 
(Hohenberg and Kohn, 1964; Kohn and Sham, 1965; Parr and Yang, 1989; Dreizler and Gross, 1990) has become one of the most successful and widespread approaches. Apart from detailed information about the electronic structure itself, the typical output of such DFT calculations, that is of relevance for the present discussion, is the energetics, e.g. total energies, as well as the forces acting on the nuclei for a given atomic configuration of a microscopically-sized system. If this energetic information is provided as function of the atomic configuration $\{ {\bf R}_I \}$, one talks about a potential energy surface (PES) $E(\{ {\bf R}_I \})$. Obviously, a (meta)stable atomic configuration corresponds to a (local) minimum of the PES. The forces acting on the given atomic configuration are just the local gradient of the PES, and the vibrational modes of a (local) minimum are given by the local PES curvature around it. Although DFT mostly does not meet the frequent demand for ``chemical accuracy'' (1 kcal/mol $\approx$\,0.04\,eV/atom) in the energetics, it is still often sufficiently accurate to allow for the aspired modeling with predictive character. In fact, we will see throughout this chapter that error cancellation at the statistical interplay level may give DFT-based approaches a much higher accuracy than may be expected on the basis of the PES alone. 

With the computed DFT forces it is possible to directly follow the motion of the atoms according to Newton's laws (Allen and Tildesley, 1997; Frenkel and Smit, 2002). With the resulting {\em ab initio molecular dynamics} (MD) (Car and Parrinello, 1985; Payne {\em et al.}, 1992; Galli and Pasquarello, 1993; Gross, 1998; Kroes, 1999) only time scales up to the order of 50 picoseconds are, however, currently accessible. Longer times may e.g. be reached by so-called accelerated MD techniques (Voter, Montalenti, and Germann, 2002), but for the desired description of a truly mesoscopic scale system which treats the statistical interplay of a large number of elementary processes over some seconds or longer, a match or combination of DFT with concepts from statistical mechanics or thermodynamics must be found. In the latter approaches, bridging of the time scale is achieved by either a suitable ``coarse-graining'' in time (to be specified below) or by only considering thermodynamically stable states.

We will discuss how such a description, appropriate for a mesoscopic scale system, can be achieved starting from electronic structure theory, as
well as ensuing concepts like {\em atomistic thermodynamics}, {\em lattice-gas Hamiltonian}, {\em equilibrium Monte Carlo simulations}, or {\em kinetic Monte Carlo simulations}. Which of these approaches (or a combination) is most suitable depends on the particular type of problem. Table~\ref{approaches} lists the different theoretical approaches and the time and length scales that they treat. While the concepts are general, we find it instructive to illustrate their power and limitations on the basis of a particular issue that is central to the field of surface-related studies including applications as important as crystal growth and heterogeneous catalysis, namely to treat the effect of a finite gas-phase. With surfaces forming the interface to the surrounding environment, a critical dependence of their properties on the species in this gas-phase, on their partial pressures and on the temperature can be intuitively expected (Zangwill, 1988; Masel, 1996). After all, we recall that for example in our oxygen-rich atmosphere, each atomic site of a close-packed crystal surface at room temperature is hit by of the order of 10$^{9}$ O$_{2}$ molecules per second. That this may have profound consequences on the surface structure and composition is already highlighted by the every-day phenomena of oxide formation, and in humid oxygen-rich environments, eventually corrosion with rust and verdigris as two visible examples (Stampfl {\em et al.}, 2002). In fact, what is typically called a stable surface structure is nothing but the statistical average over all elementary adsorption processes from, and desorption processes to, the surrounding gas-phase. If atoms or molecules of a given species adsorb more frequently from the gas-phase than they desorb to it, the species' concentration in the surface structure will be enriched with time, thus also increasing the total number of desorption processes. Eventually this total number of desorption processes will (averaged over time) equal the number of adsorption processes, the (average) surface composition and structure will remain constant, and the surface has attained its thermodynamic equilibrium with the surrounding environment. 

Within this context we may be interested in different aspects; for example, on the microscopic level, the first goal would be to separately study elementary processes such as adsorption and desorption in detail. With DFT one could e.g. address the energetics of the binding of the gas-phase species to the surface in a variety of atomic configurations (Scheffler and Stampfl, 2000),
and molecular dynamics simulations could shed light on the possibly intricate gas-surface dynamics during one individual adsorption process (Darling and Holloway, 1995; Gross, 1998; Kroes, 1999). Already the search for the most stable surface structure under given gas-phase conditions, however, requires the consideration of the interplay between the elementary processes (of at least adsorption and desorption) at the mesoscopic scale. If we are only interested in the equilibrated system, i.e. when the system has reached its thermodynamic ground state, the natural choice would then be to combine DFT data with thermodynamic concepts. How this can be done will be exemplified in the first part of this chapter. On the other hand, the processes altering the surface geometry and composition from a known initial state to the final ground state can be very slow. And coming back to the above example of oxygen-metal interaction, corrosion is a prime example, where such a kinetic hindrance significantly slows down (and practically stops) further oxidation after a passive oxide film of certain thickness has formed at the surface. In such circumstances, a thermodynamic description will not be satisfactory and one would want to follow the explicit kinetics of the surface in the given gas-phase. Then the combination of DFT with concepts from statistical mechanics explicitly treating the kinetics is required, and we will illustrate some corresponding attempts in the last section entitled ``First-principles kinetic Monte Carlo simulations''.\\

{\bf Ab initio atomistic thermodynamics} \\

Let us at first discuss the matching of electronic structure theory data with thermodynamics. Although this approach applies ``only'' to systems in equilibrium, we note that at least at not too low temperatures, a surface is likely to rapidly attain thermodynamic equilibrium with the ambient atmosphere. And even if it has not yet equilibrated, at some later stage it will have and we can nevertheless learn something by knowing about this final state. Thermodynamic considerations also have the virtue of requiring comparably less microscopic information, typically only about the minima of the PES and the local curvatures around them. As such it is often advantageous to first resort to a thermodynamic description, before embarking upon the more demanding kinetic modeling described in the last section.

The goal of the thermodynamic approach is to use the data from electronic structure theory, i.e. the information on the PES, to calculate appropriate thermodynamic potential functions like the Gibbs free energy $G$ (Kaxiras {\em et al.}, 1987; Scheffler, 1988; Scheffler and Dabrowski, 1988; Qian, Martin and Chadi, 1988). Once such a quantity is known, one is immediately in the position to evaluate macroscopic system properties. Of particular relevance for the spatial aspect of our multiscale endeavor is further that within a thermodynamic description larger systems may readily be divided into smaller subsystems that are mutually in equilibrium with each other. Each of the smaller and thus potentially simpler subsystems can then first be treated separately, and the contact between the subsystems is thereafter established by relating their corresponding thermodynamic potentials. Such a ``divide and conquer'' type of approach can be especially efficient, if infinite, but homogeneous parts of the system like bulk or surrounding gas-phase can be separated off (Wang {\em et al.}, 1998; Wang, Chaka and Scheffler, 2000; Reuter and Scheffler, 2002; Reuter and Scheffler, 2003a,b; {\L}odzianan and N{\o}rskov, 2003).\\

{\em Chemical potential plots for surface oxide formation}\\

\begin{figure}
\scalebox{0.42}{\includegraphics{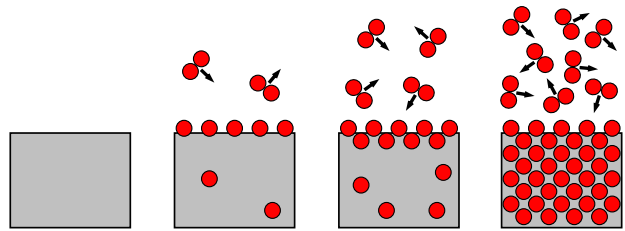}}
\caption{\label{fig_oxischematic}
Cartoon sideviews illustrating the effect of an increasingly oxygen-rich atmosphere on a metal surface. Whereas the clean surface prevails in perfect vacuum (left), finite O${}_2$ pressures in the environment also lead to an oxygen-enrichment in the solid and its surface. Apart from some bulk dissolved oxygen, frequently observed stages in this oxidation process comprise (from left to right) on-surface adsorbed O, the formation of thin (surface) oxide films, and eventually the complete transformation to an ordered bulk oxide compound.
Note, that all stages can be strongly kinetically-inhibited. It is for example, not clear whether the observation of a thin surface oxide film means that this is the stable surface composition and structure at the given gas-phase pressure and temperature, or whether the system has simply not yet attained its real equilibrium structure (possibly in form of the full bulk oxide).}
\end{figure}

How this quite general concept works and what it can contribute in practice may be illustrated with the case of oxide formation at late transition metal (TM) surfaces sketched in Figure \ref{fig_oxischematic} (Reuter and Scheffler, 2004a,b). These materials have widespread technological use, for example in the area of oxidation catalysis (Ertl, Kn\"{o}zinger and Weitkamp, 1997).
Although they are likely to form oxidic structures (i.e. ordered oxygen-metal compounds) in technologically-relevant high oxygen pressure environments, it is difficult to address this issue at the atomic scale with the corresponding experimental techniques of surface science that often require Ultra-High Vacuum (UHV) (Woodruff and Delchar, 1994). Instead of direct, so-called {\em in-situ} measurements, the surfaces are usually first exposed to a defined oxygen dosage, and the produced oxygen-enriched surface structures are then cooled down and analyzed in UHV. Due to the low temperatures, it is hoped that the surfaces do not attain their equilibrium structure in UHV during the time of the measurement, and thus provide information about the corresponding surface structure at higher oxygen pressures. This is, however, not fully certain, and it is also not guaranteed that the surface has reached its equilibrium structure during the time of oxygen exposure at first. Typically, a whole zoo of potentially kinetically-limited surface structures can be produced this way. Even though it can be academically very interesting to study all of them in detail, one would still like to have some guidance as to which of them would ultimately correspond to an equilibrium structure under which environmental conditions. Furthermore, the knowledge of a corresponding, so-called {\em surface phase diagram} as a function of in this case the temperature $T$ and oxygen pressure $p_{{\rm O}_2}$ can also provide useful information to the now surging {\em in-situ} techniques, as to which phase to expect under which environmental conditions.

The task for an {\em ab initio} atomistic thermodynamic approach would therefore be to screen a limited number of known (or otherwise provided) oxygen-containing surface structures, and evaluate which of them turns out to be the most stable one under which $(T, p_{{\rm O}_2})$ conditions (Reuter and Scheffler, 2002; Reuter and Scheffler, 2003a,b; {\L}odzianan and N{\o}rskov, 2003). Most stable translated into the thermodynamic language meaning that the corresponding structure minimizes an appropriate thermodynamic function, which would in this case be the Gibbs free energy of adsorption $\Delta G$ (Li, Stampfl, and Scheffler, 2003a,b). In other words, one has to compute $\Delta G$ as a function of the environmental variables for each structural model, and the one with the lowest $\Delta G$ is identified as most stable. Obviously this is an indirect approach, and one of the first limitations of {\em ab initio} atomistic thermodynamics studies of this kind is thus that their reliability is restricted to the structural configurations considered. If the really most stable phase is not included in the set of trial structures, the approach will not find it, although the obtained phase diagram can well give some guidance to what other structures one could and should test as well. 

What needs to be computed are all thermodynamic potentials entering into the thermodynamic function to be minimized. In the present case of the Gibbs free energy of adsorption these are for example the Gibbs free energies of bulk and surface structural models, as well as the chemical potential of the O$_2$ gas-phase. The latter may, at the accuracy level necessary for the surface phase stability issue, well be approximated by an ideal gas. The calculation of the chemical potential $\mu_{\rm O}(T,p_{{\rm O}_2})$ is then elementary and can be found in standard statistical mechanics text books (e.g. McQuarrie, 1976). Required input from a microscopic theory like DFT are properties like bond lengths and vibrational frequencies of the gas-phase species. Alternatively, the chemical potential may be directly obtained from thermochemical tables (Stull and Prophet, 1971). Compared to this, the evaluation of the Gibbs free energies of the solid bulk and surface is more involved. While in principle contributions from total energy, vibrational free energy or configurational entropy have to be calculated (Reuter and Scheffler, 2002; Reuter and Scheffler, 2003a,b), a key point to notice here is that not the absolute Gibbs free energies enter into the computation of $\Delta G$, but only the {\em difference} of the Gibbs free energies of bulk and surface. This often implies some error cancellation in the DFT total energies. It also leads to quite some (partial) cancellation in the free energy contributions like the vibrational energy. In a physical picture, it is thus not the effect of the absolute vibrations that matters for our considerations, but only the {\em changes} of vibrational modes at the surface as compared to the bulk. Under such circumstances it may result that the difference between the bulk and surface Gibbs free energies is already well approximated by the difference of their leading total energy terms, i.e. the direct output of the electronic DFT calculations (Reuter and Scheffler, 2002). Although this is of course appealing from a computational point of view, and one would always want to formulate the thermodynamic equations in a way that they contain such differences, we stress that it is not a general result and needs to be carefully checked for every specific system.

\begin{figure}
\scalebox{0.4}{\includegraphics{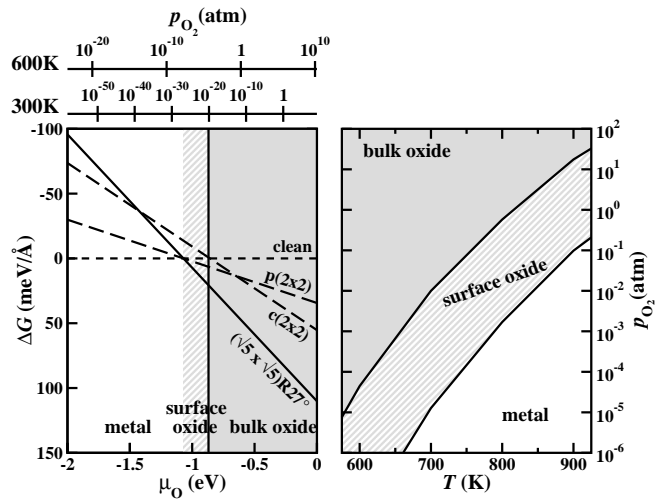}}
\caption{\label{fig_pdsurfoxide}
Left panel: Computed Gibbs free energy of adsorption $\Delta G$ for the clean Pd(100) surface and several oxygen-containing surface structures. Depending on the chemical potential $\mu_{\rm O}$ of the surrounding gas-phase, either the clean surface or a surface oxide film (labeled here according to its twodimensional periodicity as $(\sqrt{5} \times \sqrt{5})R27^{\circ}$), or the infinite PdO bulk oxide exhibit the lowest $\Delta G$ and result as the stable phase under the corresponding environmental conditions (as indicated by the different background shadings). The two structures with ordered O adlayers on the surface, $p(2 \times 2)$ and $c(2 \times 2)$, are never most stable, and their frequent observation in UHV experiments appears to be the outcome of the finite oxygen dosage and kinetic limitations at the low temperatures employed. Right panel: The stability range of the three phases, evaluated in a) as a function of $\mu_{\rm O}$, plotted directly in $(T,p_{{\rm O}_2})$-space. Note the extended stability range of the surface oxide compared to the PdO bulk oxide 
(from Reuter and Scheffler, 2004a; Lundgren {\em et al.}, 2004).}
\end{figure}

Once the Gibbs free energies of adsorption $\Delta G(T,p_{{\rm O}_2})$ are calculated for each surface structural model, they can be plotted as a function of the environmental conditions. In fact, under the imposed equilibrium the two-dimensional dependence on $T$ and $p_{{\rm O}_2}$ can be summarized into a one-dimensional dependence on the gas-phase chemical potential $\mu_{\rm O}(T,p_{{\rm O}_2})$ (Reuter and Scheffler, 2002). This is done in Figure \ref{fig_pdsurfoxide}a for the Pd(100) surface including, apart from the clean surface, a number of previously characterized oxygen-containing surface structures. These are two structures with ordered on-surface O adsorbate layers of different density [$p(2 \times 2)$ and $c(2 \times 2)$], a so-called $(\sqrt{5} \times \sqrt{5})R27^{\circ}$ surface oxide containing one layer of PdO on top of Pd(100), and finally the infinitely thick PdO bulk oxide (Todorova {\em et al.}, 2003). If we start at very low oxygen chemical potential, corresponding to a low oxygen concentration in the gas-phase, we expectedly find the clean Pd(100) surface to yield the lowest $\Delta G$ line, which in fact is used here as the reference zero. Upon increasing $\mu_{\rm O}$ in the gas-phase, the Gibbs free energies of adsorption of the other oxygen-containing surfaces decrease gradually, however, as it becomes more favorable to stabilize such structures with more and more oxygen atoms being present in the gas-phase. The more oxygen the structural models contain, the steeper the slope of their $\Delta G$ curves becomes, and above a critical $\mu_{\rm O}$ we eventually find the surface oxide to be more stable than the clean surface. Since the PdO bulk oxide contains infinitely many oxygen atoms, the slope of its $\Delta G$ line exhibits an ``infinite'' slope and cuts the other lines vertically at $\Delta \mu_{\rm O} \approx -0.8$\,eV. For any higher oxygen chemical potential in the gas-phase, the bulk PdO phase will then always result as most stable. 

With the clean surface, the surface and the bulk oxide, the thermodynamic analysis yields therefore three equilibrium phases for Pd(100) depending on the chemical potential of the O$_2$ environment. Exploiting ideal gas laws, this one dimensional dependence can be translated into the physically more intuitive dependence on temperature and oxygen pressure. For two fixed temperatures, this is also indicated by the resulting pressure scales at the top axis of Figure \ref{fig_pdsurfoxide}a. Alternatively, the stability range of the three phases can be directly plotted in $(T,p_{{\rm O}_2})$-space, as shown Figure \ref{fig_pdsurfoxide}b. Two things are worth noticing. First, in the considered thermodynamic equilibrium the two on-surface O adsorbate structures, $p(2 \times 2)$ and $c(2 \times 2)$, never correspond to a stable surface phase, suggesting that their frequent observation in UHV experiments is the mere outcome of the finite oxygen dosage or kinetic-limitations in the afore described preparation procedure. This is indeed an apparently unusual result (not yet found for other metal surfaces), and, in fact, it is not yet understood. Second, the thermodynamic stability range of the recently identified surface oxide extends well beyond the one of the common PdO bulk oxide, i.e. the surface oxide could well be present under environmental conditions where the PdO bulk oxide is known to be unstable. Also this result is somewhat unexpected, because hitherto it had been believed that it is the kinetics (not the thermodynamics) that exclusively controls the thickness of oxide films at surfaces. The additional stabilization of the $(\sqrt{5} \times \sqrt{5})R27^{\circ}$ surface oxide is attributed to the strong coupling of the ultrathin film to the Pd(100) substrate (Todorova {\em et al.}, 2003). Similar findings have recently been obtained at the Pd(111) (Lundgren {\em et al.}, 2002; Reuter and Scheffler, 2004a) and Ag(111) (Li, Stampfl and Scheffler, 2003b; Michaelides {\em et al.}, 2003) surfaces. Interestingly, the low stability of the bulk oxide phases of these more noble TMs had hitherto often been used as argument against the relevance of oxide formation in technological environments like in oxidation catalysis (Ertl, Kn\"ozinger and Weitkamp, 1997). It remains to be seen whether the surface oxide phases and their extended stability range, which have recently been intensively discussed, will change this common perception.\\

{\em Chemical potential plots for semiconductor surfaces}\\

Already in the introduction we had mentioned that the concepts discussed here are general and applicable to a wide range of problems. To illustrate this, we supplement the discussion by an example from the field of semiconductors, where the concepts of {\em ab initio} atomistic thermodynamics had in fact been developed first (Kaxiras {\em et al.}, 1987; Scheffler, 1988; Scheffler and Dabrowski, 1988; Qian, Martin and Chadi, 1988). Semiconductor surfaces exhibit complex reconstructions, i.e. surface structures that differ significantly in their atomic composition and geometry from the one that would be obtained by simply slice-cutting the bulk crystal (Zangwill, 1988). Knowledge of the surface atomic structure is, on the other hand, a prerequisite to understand and control the surface or interface electronic properties, as well as the detailed growth characteristics. While the number of possible configurations with complex surface unit-cell reconstructions is already large, searching for possible structural models becomes even more involved for surfaces of compound semiconductors. In order to minimize the number of dangling bonds, the surface may exchange atoms with the surrounding gas-phase, which in molecular beam epitaxy (MBE) growth is composed of the substrate species at elevated temperatures and varying partial pressures. As a consequence of the interaction with this gas-phase, the surface stoichiometry may be altered and surface atoms be displaced to assume a more favorable bonding geometry. The resulting surface structure depends thus on the environment, and atomistic thermodynamics may again be employed to compare the stability of existing (or newly suggested) structural models as a function of the conditions in the surrounding gas-phase. The thermodynamic quantity that is minimized by the most stable structure is in this case the surface free energy, which in turn depends on the Gibbs free energies of the bulk and surface of the compound, as well as on the chemical potentials in the gas-phase. The procedure of evaluating these quantities goes exactly along the lines described above, where in addition, one frequently assumes the surface fringe not only to be in thermodynamic equilibrium with the surrounding gas-phase, but also with the underlying compound bulk (Reuter and Scheffler, 2002). With this additional constraint, the dependence of the surface structure and composition on the environment can, even for the two component gas-phase in MBE, be discussed as a function of the chemical potential of only one of the compound species alone.

\begin{figure}
\scalebox{0.43}{\includegraphics{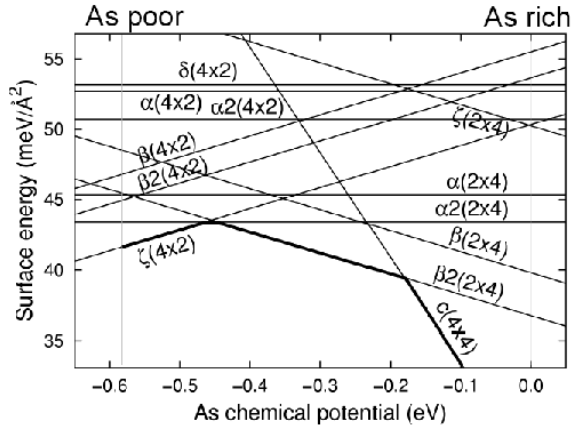}}
\caption{\label{fig_sunghoon}
Surface energies for GaAs(001) terminations as a function of the As chemical potential, $\mu_{\rm As}$. The thermodynamically allowed range of $\mu_{\rm As}$ is bounded by the formation of Ga droplets at the surface (As-poor limit at -0.58\,eV) and the condensation of arsenic at the surface (As-rich limit at 0.00\,eV). The $\zeta(4\times2)$ geometry is significantly lower in energy than the previously proposed $\beta2(4 \times 2)$ model for the $c(8 \times 2)$ surface reconstruction observed under As-poor growth conditions (from Lee, Moritz and Scheffler, 2000).}
\end{figure}

Figure \ref{fig_sunghoon} shows as an example the dependence on the As content in the gas-phase for a number of surface structural models of the GaAs(001) surface. A reasonable lower limit for this content is given, when there is so little As$_2$ in the gas-phase that it becomes thermodynamically more favorable for the arsenic to leave the compound. The resulting GaAs decomposition and formation of Ga droplets at the surface denotes the lower limit of As chemical potentials considered (As-poor limit), while the condensation of arsenic on the surface forms an appropriate upper bound (As-rich limit). Depending on the As to Ga stoichiometry at the surface, the surface free energies of the individual models have either a positive slope (As-poor terminations), a negative slope (As-rich terminations) or remain constant (stoichiometric termination). While the detailed atomic geometries behind the considered models in Figure \ref{fig_sunghoon} are not relevant here, most of them may roughly be characterized as different ways of forming dimers at the surface in order to reduce the number of dangling orbitals (Duke, 1996). In fact, it is this general ``rule'' of dangling bond minimization by dimer formation that has hitherto mainly served as inspiration in the creation of new structural models for the (001) surfaces of III-V zinc-blende semiconductors, thereby leading to some prejudice in the type of structures considered. In contrast, the at first theoretically proposed so-called $\zeta(4 \times 2)$ structure is actuated by the filling of all As dangling orbitals and emptying of all Ga dangling orbitals, as well as a favorable electrostatic (Ewald) interaction between the surface atoms (Lee, Moritz and Scheffler, 2000). The virtue of the atomistic thermodynamic approach is now that such a new structural model can be directly compared in its stability against all existing ones. And indeed, the $\zeta(4 \times 2)$ phase was found to be significantly more stable than all previously (experimentally) proposed reconstructions at low As pressure. 

Returning to the methodological discussion, the results shown in Figures  \ref{fig_pdsurfoxide} and \ref{fig_sunghoon} nicely summarize the contribution that can be made by such {\em ab initio} atomistic thermodynamics considerations. On the other hand, they also highlight the limitations. Most prominently, one has to be aware that the reliability is restricted to the number of considered configurations, or in other words that only the stability of exactly those structures plugged in can be compared. Had for example the surface oxide structure not been known and not explicitly been considered in Figure \ref{fig_pdsurfoxide}, the $p(2 \times 2)$ adlayer structure would have yielded the lowest Gibbs free energy of adsorption in a range of $\mu_{\rm O}$ intermediate to the stability ranges of the clean surface and the bulk oxide, changing the resulting surface phase diagram accordingly. Alternatively, it is at present not completely clear, whether the $(\sqrt{5} \times \sqrt{5})R27^{\circ}$ structure is really the {\em only} surface oxide on Pd(100). If another yet unknown surface oxide exists and exhibits a sufficiently low $\Delta G$ for some oxygen chemical potential, it will similarly affect the surface phase diagram, as would another novel and hitherto unconsidered surface reconstruction with sufficiently low surface free energy in the GaAs example. As such, appropriate care should be in place when addressing systems where only limited information about surface structures is available. With this in mind, even in such systems the atomistic thermodynamics approach can still be a particularly valuable tool though, since it allows for example to rapidly compare the stability of newly devised structural models against existing ones.

In the section entitled ``{\em Ab initio} lattice-gas Hamiltonian'' we will discuss an approach that is able to overcome this limitation. This comes unfortunately at a significantly higher computational demand, so that it has up to now only be used to study simple adsorption layers on surfaces. This will then also provide more detailed insight into the transitions between stable phases. In Figures \ref{fig_pdsurfoxide} and \ref{fig_sunghoon} the transitions are simply drawn abrupt, and no reference is made to the finite phase coexistence regions that should occur at finite temperatures, i.e. regions in which with changing pressure or temperature one phase gradually becomes populated and the other one depopulated. That this is not the case in the discussed examples has to do with that the configurational entropy contribution to the Gibbs free energy of the surface phases has been neglected in the corresponding studies (Reuter and Scheffler, 2003b). For the well-ordered surface structural models considered, this contribution is indeed small and will affect only a small region close to the phase boundaries. The width of this affected phase coexistence region can even be estimated, but if more detailed insight into this very region is desired, or if disorder becomes more important e.g. at more elevated temperatures, then an explicit calculation of the configurational entropy contribution will become necessary. For this, equilibrium Monte Carlo simulations as described below are the method of choice, but before we turn to them there is yet another twist to chemical potential plots that deserves mentioning.\\

{\em ``Constrained equilibrium''}\\

Although a thermodynamic approach can strictly describe only the situation where the surface is in equilibrium with the surrounding gas-phase, the idea is that it can still give some insight when the system is {\em close} to thermodynamic equilibrium, or even when it is only close to thermodynamic equilibrium with some of the present gas-phase species (Reuter and Scheffler, 2003a). For such situations it can be useful to consider ``constrained equilibria'', and one would expect to get some ideas as to where in $(T,p)$-space thermodynamic phases may still exist, but also to identify those regions where kinetics may control the material function. 

We will discuss heterogeneous catalysis as a prominent example. Here, a constant stream of reactants is fed over the catalyst surface and the formed products are rapidly carried away. If we take the CO oxidation reaction to further specify our example, the surface would be exposed to an environment composed of O${}_2$ and CO molecules, while the produced CO${}_2$ desorbs from the catalyst surface at the technologically employed temperatures and is then transported away. Neglecting the presence of the CO$_2$, one could therefore model the effect of an O${}_2$/CO gas-phase on the surface, in order to get some first ideas of the structure and composition of the catalyst under steady-state operation conditions. Under the assumption that the adsorption and desorption processes of the reactants occur much faster than the CO$_2$ formation reaction, the latter would not significantly disturb the average surface population, i.e. the surface could be close to maintaining its equilibrium with the reactant gas-phase. If at all, this equilibrium holds, however, only with each gas-phase species separately. Were the latter fully equilibrated among each other, too, only the products would be present under all environmental conditions of interest. It is in fact particularly the high free energy barrier for the direct gas-phase reaction that prevents such an equilibration on a reasonable time scale, and necessitates the use of a catalyst in the first place. 

The situation that is correspondingly modeled in an atomistic thermodynamics approach to heterogeneous catalysis is thus a surface in ``constrained equilibrium'' with {\em independent} reservoirs representing all {\em reactant} gas-phase species, namely O$_2$ and CO in the present example (Reuter and Scheffler, 2003a). It should immediately be stressed though, that such a setup should only be viewed as a thought construct to get a first idea about the catalyst' surface structure in a high-pressure environment. Whereas we could write before that the surface will sooner or later necessarily equilibrate with the gas-phase in the case of a pure O$_2$ atmosphere, this must no longer be the case for a ``constrained equilibrium''. The on-going catalytic reaction at the surface consumes adsorbed reactant species, i.e. it continuously drives the surface populations away from their equilibrium value, and even more so in the interesting regions of high catalytic activity.

\begin{figure}
\scalebox{0.49}{\includegraphics{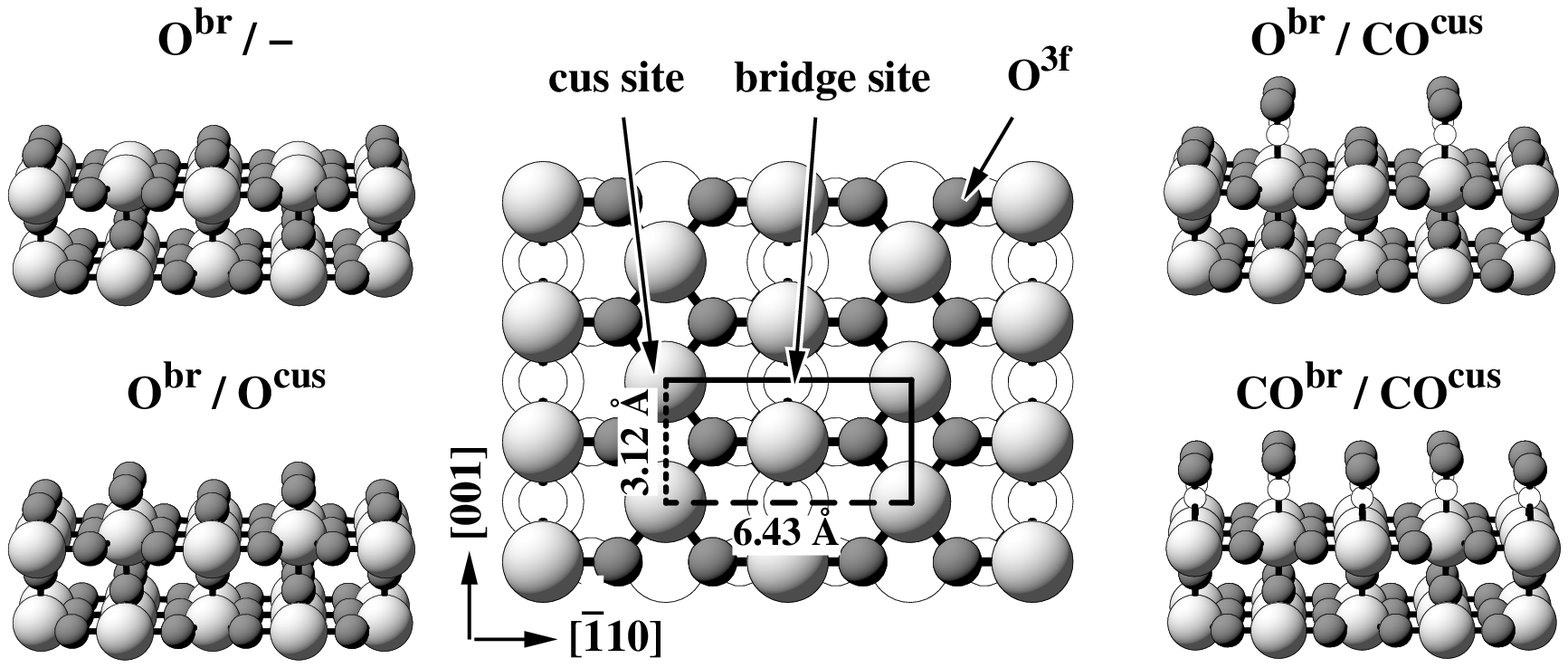}}
\scalebox{0.44}{\includegraphics{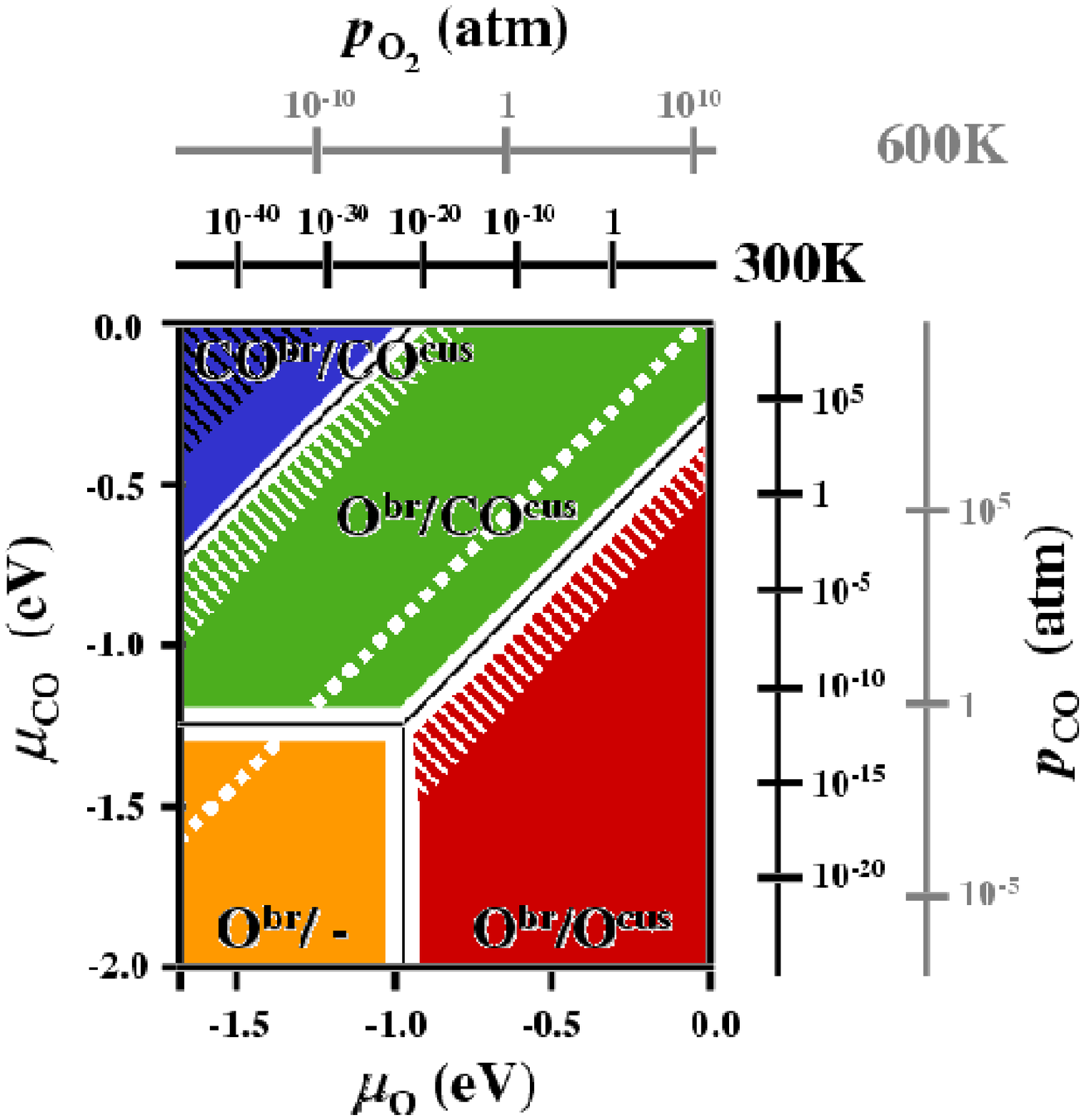}}
\caption{\label{fig_oruphasediag}
Top panel: Top view of the RuO$_2$(110) oxide surface explaining the location of the two prominent adsorption sites (coordinatively unsaturated, cus, and bridge, br). Also shown are side views of the four stable phases present in the phase diagram shown below (Ru = light large spheres, O = dark medium spheres, C = white small spheres). Bottom panel: Surface phase diagram for RuO$_2$(110) in ``constrained equilibrium'' with an oxygen and CO environment. Depending on the gas-phase chemical potentials ($\mu_{\rm O}, \mu_{\rm CO}$), br and cus sites are either occupied by O or CO, or empty (--), yielding a total of four different surface phases. For $T=300$\,K and $T=600$\,K, this dependence is also given in the corresponding pressure scales. Regions that are expected to be particularly strongly affected by phase coexistence or kinetics are marked by white hatching (see text). Note that conditions representative for technological CO oxidation catalysis (ambient pressures, 300-600\,K) fall exactly into one of these ranges (from Reuter and Scheffler, 2003a,b).}
\end{figure}

That the ``constrained equilibrium'' concept can still yield valuable insight is nicely exemplified for the CO oxidation over a Ru catalyst (Engel and Ertl, 1982). For ruthenium the afore described tendency to oxidize under oxygen-rich environmental conditions is much more pronounced than for the above discussed nobler metals Pd and Ag (Reuter and Scheffler, 2004a). While for the latter the relevance of (surface) oxide formation under the conditions of technological oxidation catalysis is still under discussion (Reuter and Scheffler, 2004a; Li, Stampfl and Scheffler, 2003b; Michaelides {\em et al.} 2003; Hendriksen, Bobaru and Frenken, 2003), it is by now established that a film of bulk-like oxide forms on the Ru(0001) model catalyst during high-pressure CO oxidation, and that this RuO$_2$(110) is the active surface for the reaction (Over and Muhler, 2003). When evaluating its surface structure in ``constrained equilibrium'' with an O$_2$ and CO environment, four different stable phases result depending on the gas-phase conditions that are now described by the chemical potentials of both reactants, cf. Figure \ref{fig_oruphasediag}. The phases differ from each other in the occupation of two prominent adsorption site types exhibited by this surface, called bridge (br) and coordinatively unsaturated (cus) sites. At very low $\mu_{\rm CO}$, i.e. a very low CO concentration in the gas-phase, either only the bridge, or bridge and cus sites are occupied by oxygen depending on the O$_2$ pressure. Under increased CO concentration in the gas-phase, both the corresponding O$^{\rm br}/-$ and the O$^{\rm br}$/O$^{\rm cus}$ phase have to compete with CO that would also like to adsorb at the cus sites. And eventually the O$^{\rm br}$/CO$^{\rm cus}$ phase develops. Finally, under very reducing gas-phase conditions with a lot of CO and essentially no oxygen, a completely CO covered surface results (CO$^{\rm br}$/CO$^{\rm cus}$). Under these conditions the RuO$_2$(110) surface can at best be metastable, however, as above the white-dotted line in Figure \ref{fig_oruphasediag} the RuO$_2$ bulk oxide is already unstable against CO-induced decomposition.

With the already described difficulty of operating the atomic-resolution techniques of surface science at high pressures, the possibility of reliably bridging the so-called pressure gap is of key interest in heterogeneous catalysis research (Ertl, Kn\"ozinger, and Weitkamp, 1997; Engel and Ertl, 1982; Ertl, 2002). The hope is that the atomic-scale understanding gained in experiments with some suitably chosen low pressure conditions would also be representative of the technological ambient pressure situation. Surface phase diagrams like the one shown in Figure \ref{fig_oruphasediag} could give some valuable guidance in this endeavor. If the $(T,p_{{\rm O}_2},p_{\rm CO})$  conditions of the low pressure experiment are chosen such that they lie within the stability region of the same surface phase as at high-pressures, the same surface structure and composition will be present and scalable results may be expected. If, however, temperature and pressure are varied in such a way, that one crosses from one stability region to another one, different surfaces are exposed and there is no reason to hope for comparable functionality. This would e.g. also hold for a naive bridging of the pressure gap by simply maintaining a constant partial pressure ratio. 

In fact, the comparability holds not only within the regions of the stable phases themselves, but with the same argument also for the phase coexistence regions along the phase boundaries. The extent of these configurational entropy induced phase coexistence regions has been indicated in Figure \ref{fig_oruphasediag} by white regions. Although as already discussed, the above mentioned approach gives no insight into the detailed surface structure under these conditions, pronounced fluctuations due to an enhanced dynamics of the involved elementary processes can generally be expected due to the vicinity of a phase transition. Since catalytic activity is based on the same dynamics, these regions are therefore likely candidates for efficient catalyst functionality (Reuter and Scheffler, 2003a). And indeed, very high and comparable reaction rates have recently been noticed for different environmental conditions that all lie close to the white region between the O$^{\rm br}$/O$^{\rm cus}$ and O$^{\rm br}$/CO$^{\rm cus}$ phases. It must be stressed, however, that exactly in this region of high catalytic activity one would similarly expect the break-down of the ``constrained equilibrium'' assumption of a negligible effect of the on-going reaction on the average surface structure and stoichiometry. At least everywhere in the corresponding hatched regions in Figure \ref{fig_oruphasediag} such kinetic effects will lead to significant deviations from the surface phases obtained within the approach described above, even at ``infinite'' times after steady-state has been reached. Atomistic thermodynamics may therefore be employed to {\em identify} interesting regions in phase space. Their surface coverage and structure, i.e. the very dynamic behavior, must then however be modeled by statistical mechanics explicitly accounting for the kinetics, and the corresponding kinetic Monte Carlo simulations will be discussed towards the end of the chapter.\\

{\em Ab initio lattice-gas Hamiltonian} \\

The predictive power of the approach discussed in the previous sections extends only to the structures that are directly considered, i.e., it cannot predict the existence of unanticipated geometries or stoichiometries. It also does not explicitly describe coexistence phases or order-disorder transitions as configurational entropy is (typically) not included. To overcome both of these limitations, a proper sampling of the whole configuration space must be achieved, instead of considering only a set of structural models. Modern statistical mechanical methods like Monte Carlo (MC) simulations are particularly designed to efficiently fulfill this purpose (Frenkel and Smit, 2002; Landau and Binder, 2002). The straightforward matching with electronic structure theories would thus be to determine with DFT the energetics of all system configurations generated in the course of the statistical simulation. Unfortunately, this direct linking is currently and also in the foreseeable future computationally unfeasible. The exceedingly large configuration spaces of most materials science problems require a prohibitively large number of free energy evaluations (which can easily go beyond 10$^6$ for moderately complex systems). Furthermore, also disordered configurations must be evaluated, which in turn is not easy to achieve with the typical periodic boundary conditions of DFT supercell calculations (Scheffler and Stampfl, 2000).

\begin{figure}
\scalebox{0.33}{\includegraphics{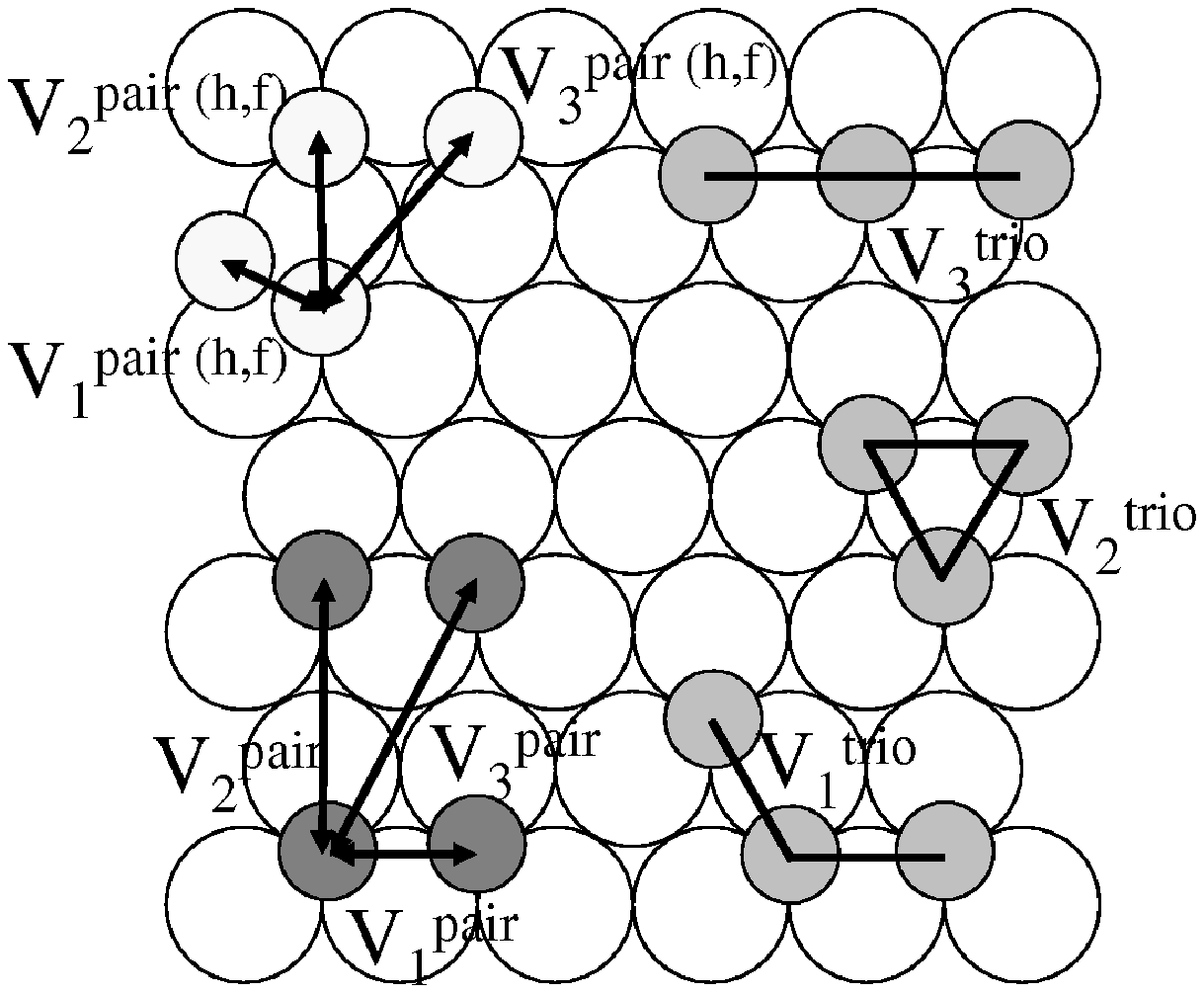}}\scalebox{0.33}{\includegraphics{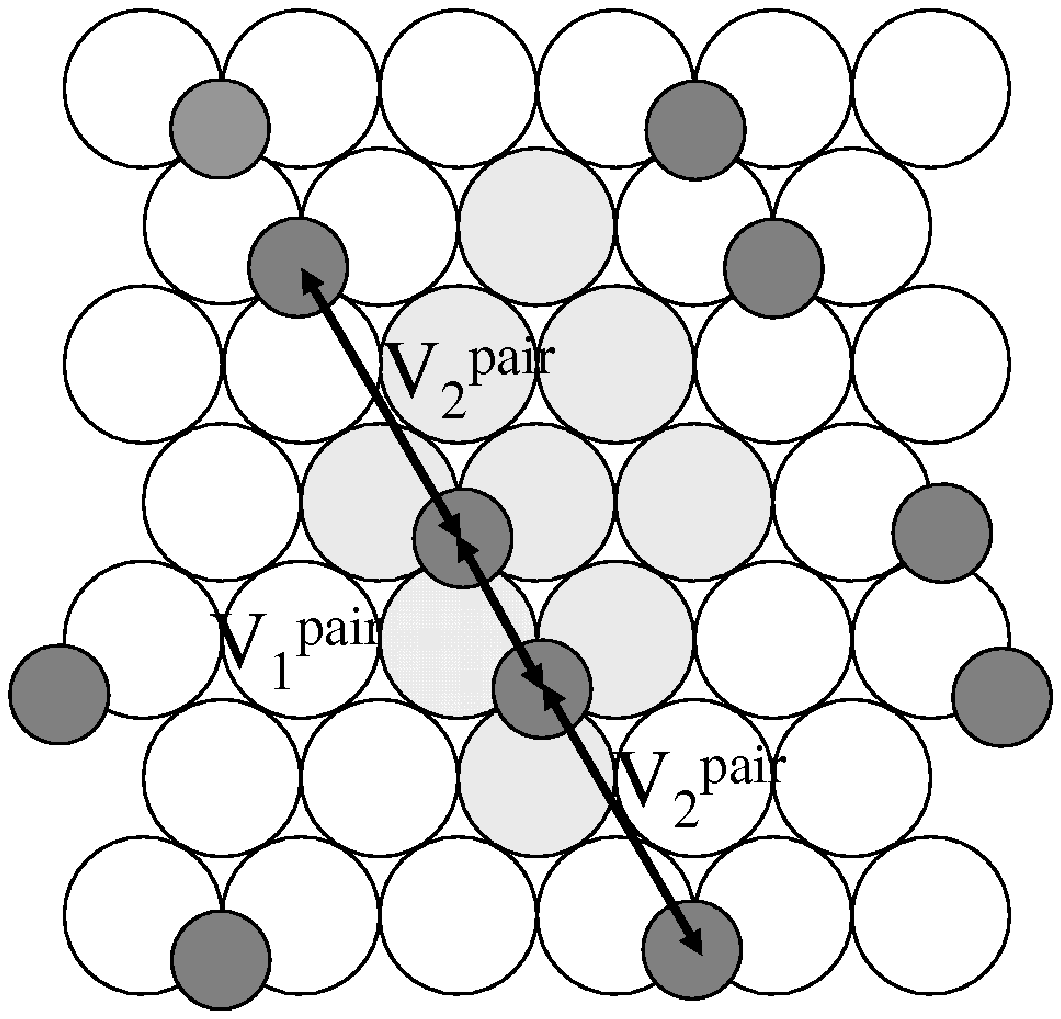}}
\caption{Left: Illustration of some types of lateral interactions for the case of a two-dimensional adsorbate layer (small dark spheres) that can occupy the two distinct threefold hollow sites of a (111) close-packed surface. $V_n^{\rm pair}$\,$(n=1,2,3)$ are two-body (or pair) interactions at first, second and third nearest neighbor distances of like hollow sites (i.e. fcc-fcc or hcp-hcp). $V_n^{\rm trio}$\,$(n=1,2,3)$ are the three possible three-body (or trio) interactions between three atoms in like nearest neighbor hollow sites, and $V_n^{\rm pair (h,f)}$\,$(n=1,2,3)$ represent pair interactions between atoms that occupy unlike hollow sites (i.e. one in fcc and the other in hcp or vice versa). Right: Example of an adsorbate arrangement from which an expression can be obtained for use in solving for interaction parameters. The $(3 \times 3)$ periodic surface unit-cell is indicated by the large darker spheres. The arrows indicate interactions between the adatoms. Apart from the obvious first 
nearest-neighbor interactions (short arrows), also third nearest-neighbor two-body interactions (long arrows) exist, due to the periodic images outside of the unit-cell.}
\label{fig_latgas}
\end{figure}

With the direct matching impossible, an efficient alternative is to map the real system somehow onto a simpler, typically discretized model system, the Hamiltonian of which is sufficiently fast to evaluate. This then enables us to evaluate the extensive number of free energies required by the statistical mechanics. Obvious uncertainties of this approach are how appropriate the model system represents the real system, and how its parameters can be determined from the first-principles calculations. The advantage, on the other hand, is that such a detour via an appropriate (``coarse-grained'') model system often provides deeper insight and understanding of the ruling mechanisms. If the considered problem can be described by a lattice defining the possible sites for the species in the system, a prominent example for such a mapping approach is given by the concept of a lattice-gas Hamiltonian (LGH) [or in other languages, an ``Ising-type model'' (de Fontaine, 1994) or a ``cluster-expansion'' (Sanchez, Ducastelle and Gratias, 1984; Zunger, 1994)]. Here, any system state is defined by the occupation of the sites in the lattice and the total energy of any configuration is expanded into a sum of discrete interactions between these lattice sites. For a one component system with only one site type, the LGH would then for example read (with obvious generalizations to multi-component, multi-site systems):

\begin{displaymath}
H  =  F \sum_{i}n_{i} \;+ \sum_{m=1}^{\rm pair}V_{m}^{\rm pair} 
\sum_{(ij)_m}n_{i}n_{j} \;+
\end{displaymath}

\begin{equation}
\sum_{m=1}^{\rm trio}V_{m}^{\rm trio} \sum_{(ijk)_{m}} n_{i}n_{j}n_{k}
+ \ldots \quad ,
\label{eq-latgas}
\end{equation}
where the site occupation numbers $n_{l}$ = 0 or 1 depend on whether site $l$ in the lattice is empty or occupied, and $F$ is the free energy of an isolated species at this lattice site, including static and vibrational contributions. $V_{m}^{\rm pair}$ are the two-body (or pair) interaction energies between species at $m$th nearest neighbor sites and $V^{\rm trio}_{m}$ is the energy due to three-body (or trio) interactions. Formally, higher and higher order interaction terms (quattro, quinto,...) would follow in this infinite expansion. In practice, the series must obviously (and can) be truncated after a finite number of terms though. Figure~\ref{fig_latgas} illustrates some of these interactions for the case of a two-dimensional adsorbate layer that can occupy the two distinct threefold hollow sites of a (111) close-packed surface. In particular, the pair interactions up to third nearest neighbor between like and unlike hollow sites are shown, as well as three possible trio interactions between adsorbates in like sites. 

It is apparent that such a LGH is very general. The Hamiltonian can be equally well evaluated for any lattice occupation, be it dense or sparse, periodic or disordered. And in all cases it merely comprises performing an algebraic sum over a finite number of terms, i.e. it is computationally very fast. The disadvantage is, on the other hand, that for more complex systems with multiple sites and several species, the number of interaction terms in the expansion increases rapidly. Precisely which of these (far-reaching or multi-body) interaction terms need to be considered, i.e. where the sum in eq. (1) may be truncated, and how the interaction energies in these terms may be determined, is the really sensitive part of such a lattice-gas Hamiltonian approach that must be carefully checked.

The methodology in itself is not new, and traditionally the interatomic interactions have often been {\em assumed} to be just pairwise additive (i.e. higher order terms beyond pair interactions were neglected); the interaction energies were then obtained by simply fitting to experimental data (see e.g. Piercy, De'Bell, and Pfn\"{u}r, 1992; Xiong {\em et al.}, 1997; Zhdanov and Kasemo, 1998). This procedure obviously results in ``effective parameters'' with an unsure microscopic basis, ``hiding'' or ``masking'' the effect and possible importance of three-body (trio) and higher-order interactions. This has the consequence that while the Hamiltonian may be able to reproduce certain specific experimental data to which the parameters were fitted, it is questionable and unlikely that it will be general and transferable to calculations of other properties of the system. Indeed, the decisive contribution to the observed behavior of adparticles by higher-order, many-atom interactions has in the meanwhile been pointed out by a number of studies (see e.g. Koh and Ehrlich, 1999; \"{O}sterlund {\em et al.}, 1999; Payne {\em et al.}, 1999).

As an alternative to this empirical procedure, the lateral interactions between the particles in the lattice can be deduced from detailed DFT calculations, and it is this approach in combination with the statistical mechanics methods that is of interest for this chapter. The straightforward way to do this is obviously to directly compute these interactions as a difference of calculations, in which once the involved species are only separately present at the corresponding lattice sites, and once all at the same time. For the example of a pair interaction between two adsorbates at a surface, this would translate into two DFT calculations where only either one of the adsorbates sits at its lattice site, and one calculation where both are present simultaneously. Unfortunately, this type of approach is hard to combine with the periodic boundary conditions that are typically required to describe the electronic structure of solids and surfaces (Scheffler and Stampfl, 2000). In order to avoid interactions with the periodic images of the considered lattice species, huge (actually often prohibitively large) supercells would be required. A more efficient and intelligent way of addressing the problem is instead to specifically exploit the interaction with the periodic images. For this, different configurations in various (feasible) supercells are computed with DFT, and the obtained energies expressed in terms of the corresponding interatomic interactions. Figure \ref{fig_latgas} illustrates this for the case of two adsorbed atoms in a laterally periodic surface unit-cell. Due to this periodicity, each atom has images in the neighboring cells. Because of these images, each of the atoms in the unit-cell experiences not only the obvious pair interaction at the first neighbor distance, but also a pair interaction at the third neighbor distance (neglecting higher pairwise or multi-body interactions for the moment). The computed DFT binding energy for this configuration $i$ can therefore be written as $E_{\rm DFT}^{(3\times3),i} = 2E + 2 V_{1}^{\rm pair} + 2 V_{3}^{\rm pair}$. Doing this for a set of different configurations thus generates a system of linear equations that can be solved for the interaction energies either by direct inversion (or by fitting techniques, if more configurations than interaction parameters were determined).

The crucial aspect in this procedure is the number and type of interactions to include in the LGH expansion, and the number and type of configurations that are computed to determine them. We note that there is no {\em a priori} way to know at how many, and what type of, interactions to terminate the expansion. While there are some attempts to automatize this procedure (van de Walle and Ceder, 2002), it is probably fair to say that the actual implementation remains to date a delicate task. Some guidelines to judge on the convergence of the constructed Hamiltonian include its ability to predict the energies of a number of DFT-computed configurations that were not employed in the fit, or that it reproduces the correct lowest-energy configurations at $T=0$\,K (so-called ``ground-state line''; Zunger, 1994).\\

{\em Equilibrium Monte Carlo simulations} \\

Once an accurate lattice-gas Hamiltonian has been constructed, one has at hand a very fast and flexible tool to provide the energies of arbitrary system configurations. This may in turn be used for Monte Carlo simulations to obtain a good sampling of the available configuration space, i.e. to determine the partition function of the system. An important aspect of modern MC techniques is that this sampling is done very efficiently by concentrating on those parts of the configuration space that contribute significantly to the latter. The Metropolis algorithm (Metropolis {\em et al.}, 1953), as a famous example of such so-called importance sampling schemes, proceeds therefore by generating at random new system configurations. If the new configuration exhibits a lower energy than the previous one, it is automatically ``accepted'' to a gradually built-up sequence of configurations. And even if the configuration has a higher energy, it still has an appropriately Boltzmann weighted probability to make it to the sequence. Otherwise it is ``rejected'' and the last configuration copied anew to the sequence. This way, the algorithm preferentially samples low energy configurations, which contribute most to the partition function. The acceptance criteria of the Metropolis and of other importance sampling schemes are furthermore designed in such a way, that they fulfill detailed balance. This means that the forward probability of accepting a new configuration $j$ from state $i$ is related to the backward probability of accepting configuration $i$ from state $j$ by the free energy difference of both configurations. Taking averages of system observables over the thus generated configuration sequences yields then their correct thermodynamic average for the considered ensemble. Technical issues regard finally how new trial configurations are generated, or how long and in what system size the simulation must be run in order to obtain good statistical averages (Frenkel and Smit, 2002; Landau and Binder, 2002).

The kind of insights that can be gained by such a first-principles LGH + MC approach is nicely exemplified by the problem of on-surface adsorption at a close-packed surface, when the latter is in equilibrium with a surrounding gas-phase. If this environment consists of oxygen, this would e.g. contribute to the understanding of one of the early oxidation stages sketched in Figure \ref{fig_oxischematic}. What would be of interest is for instance to know how much oxygen is adsorbed at the surface given a certain temperature and pressure in the gas-phase, and whether the adsorbate forms ordered or disordered phases. As outlined above, the approach proceeds by first determining a LGH from a number of DFT-computed ordered adsorbate configurations. This is followed by grand-canonical MC simulations, in which new trial system configurations are generated by randomly adding or removing adsorbates from the lattice positions and where the energies of these configurations are provided by the LGH. Evaluating appropriate order parameters that check on prevailing lateral periodicities in the generated sequence of configurations, one may finally plot the phase diagram, i.e. what phase exists under which $(T,p)$-conditions (or equivalently $(T,\mu)$-conditions) in the gas-phase. 

\begin{figure}
\scalebox{0.50}{\includegraphics{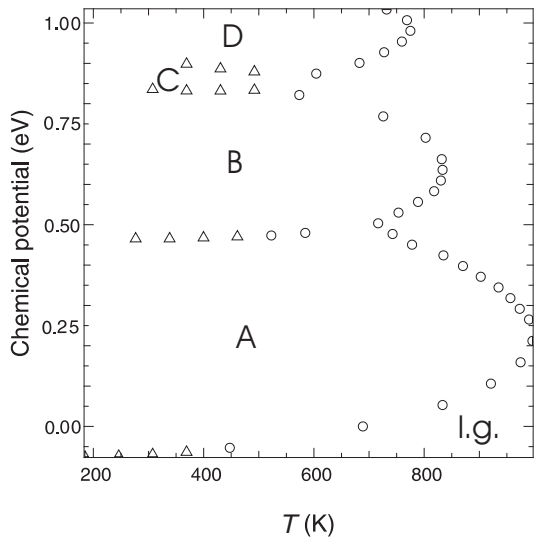}}
\caption{Phase diagram for O/Ru(0001) as obtained using the {\em ab initio} lattice-gas Hamiltonian approach in combination with MC calculations. 
The triangles indicate first order transitions and the circles second order transitions. The identified ordered structures are labeled as: $(2\times 2)$-O (A), $(2\times 1)$-O (B), $(\sqrt{3}\times \sqrt{3})R30^{\circ}$
(C), $(2\times 2)$-3O (D), and disordered lattice-gas (l.g.).
(From McEwen, Payne and Stampfl, 2002).}
\label{fig_orumcphasediag}
\end{figure}

\begin{figure}
\scalebox{1.02}{\includegraphics{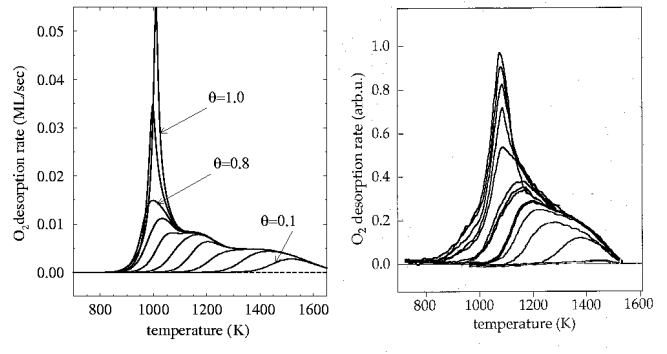}}
\caption{Theoretical (left panel) and experimental (right panel) temperature programmed desorption curves. Each curve shows the rate of oxygen molecules that desorb from the Ru(0001) surface as a function of temperature, when the system is prepared with a given initial oxygen coverage $\theta$ ranging from 0.1 to 1 monolayer. The first-principles LGH employed in the calculations is exactly the same as the one underlying the phase diagram of Figure \ref{fig_orumcphasediag} (from Stampfl {\em et al.}, 1999a,b).}
\label{fig_tpd}
\end{figure}

The result of one of the first studies of this kind is shown in Figure \ref{fig_orumcphasediag} for the system O/Ru(0001). The employed lattice-gas Hamiltonian comprised two types of adsorption sites, namely the hcp and fcc hollows, lateral pair interactions up to third neighbor and three types of trio interactions between like and unlike sites, thus amounting to a total of fifteen independent interaction parameters. At low temperature, the simulations yield a number of ordered phases corresponding to different periodicities and oxygen coverages. Two of these ordered phases had already been reported experimentally at the time the work was carried out. The prediction of two new (higher coverage) periodic structures, namely a 3/4 and a 1 monolayer phase, has in the meanwhile been confirmed by various experimental studies. This example thus demonstrates the predictive nature of the first-principles approach, and the stimulating and synergetic interplay between theory and experiment. It is also worth pointing out that these new phases and their coexistence in certain coverage regions were not obtained in early MC calculations of this system based on an empirical LGH, which was determined by simply fitting a minimal number of pair interactions to the then available experimental phase diagram (Piercy, De'Bell and Pfn\"ur, 1992). We also like to stress the superior {\em transferability} of the first-principles interaction parameters. As an example we name simulations of temperature programmed desorption (TPD) spectra, which can among other possibilities be obtained by combining the LGH with a transfer-matrix approach and kinetic rate equations (Kreuzer and Payne, 2000). Figure \ref{fig_tpd} shows the result obtained with exactly the same LGH that also underlies the phase diagram of Figure \ref{fig_orumcphasediag}. Although empirical fits of TPD spectra may give better agreement between calculated and experimental results, we note that the agreement visible in Figure \ref{fig_tpd} is in fact quite good. The advantage, on the other hand, is that no empirical parameters were used in the LGH, which allows to unambiguously trace back the TPD features to lateral interactions with well defined microscopic meaning.

The results summarized in Figure \ref{fig_orumcphasediag} serve also quite well to illustrate the already mentioned differences between the initially described chemical potential plots and the LGH + MC method. In the first approach the stability of a fixed set of configurations is compared in order to arrive at the phase diagram. For the O/Ru(0001) system, the likely choice at the time would have been just the two experimentally known ordered phases, O$(2 \times 2)$ and O$(2 \times 1)$. The stability region of the prior phase, bounded at lower chemical potentials by the clean surface and at higher chemical potentials by the O$(2 \times 1)$ phase, then comes out just as much as in Figure \ref{fig_orumcphasediag}. This stability range will be independent of temperature, however, i.e. there is no order-disorder transition at higher temperature due to the neglect of configurational entropy. More importantly, since the two new higher-coverage phases would not have been explicitly considered, the stability of the O$(2 \times 1)$ phase would falsely extend over the whole higher chemical potential range. While this emphasizes the deeper insight and increased predictive power that is achieved by the proper sampling of configuration space in the LGH + MC technique, one must also recognize that the computational cost of the latter is significantly higher. It is in particular straightforward to directly compare the stability of qualitatively different geometries like the on-surface adsorption and the surface oxide phases in Figure \ref{fig_pdsurfoxide} in a chemical potential plot. Setting up a lattice-gas Hamiltonian that would equally describe both systems, on the other hand, is far from trival. Even if it were feasible to find a generalized lattice that would be able to treat all system states, disentangling and determining the manifold of interaction energies in such a lattice will be very involved. The required discretization of the real system, i.e. the mapping onto a lattice, is therefore to date the major limitation of the LGH + MC technique -- be it applied to two-dimensional pure surface systems or even worse to three-dimensional problems addressing a surface fringe of finite width. Still, it is also precisely this mapping and the resulting very fast analysis of the properties of the LGH that allows for an extensive and reliable sampling of the configuration space of complex systems that is hitherto unparalleled by other approaches.

Having highlighted the importance of this sampling for the determination of unanticipated new ordered phases at lower temperatures, the final example in this section illustrates specifically the decisive role it plays also for the simulation and understanding of order-disorder transitions at elevated temperatures. A particularly intriguing transition of this kind is observed for Na on Al(001). The interest in such alkali metal adsorption systems has been intense, especially since in the early 1990's it was found (first for Na on Al(111) and then on Al(100)) that the alkali metal atoms may kick-out surface Al atoms and adsorb substitutionally. This was in sharp contrast to the generally accepted understanding of the time, which was that alkali-metal atoms adsorb in the highest coordinated on-surface hollow site, and cause little disturbance to a close-packed metal surface (Stampfl and Scheffler, 1995; Adams, 1996; Diehl and Mc Grath, 1997). For the specific system Na on Al(001) at a coverage of 0.2\,monolayers, a reversible phase transition is observed in low energy electron diffraction experiments at $T= 240$\,K. Below this temperature, an ordered $(\sqrt{5}\times \sqrt{5})R27^{\circ}$ structure forms, where the Na atoms occupy surface substitutional sites. At temperatures above 240\,K on the other hand, the Na atoms, still in the same substitutional sites, form a disordered arrangement in the surface.

\begin{figure}
\scalebox{0.45}{\includegraphics{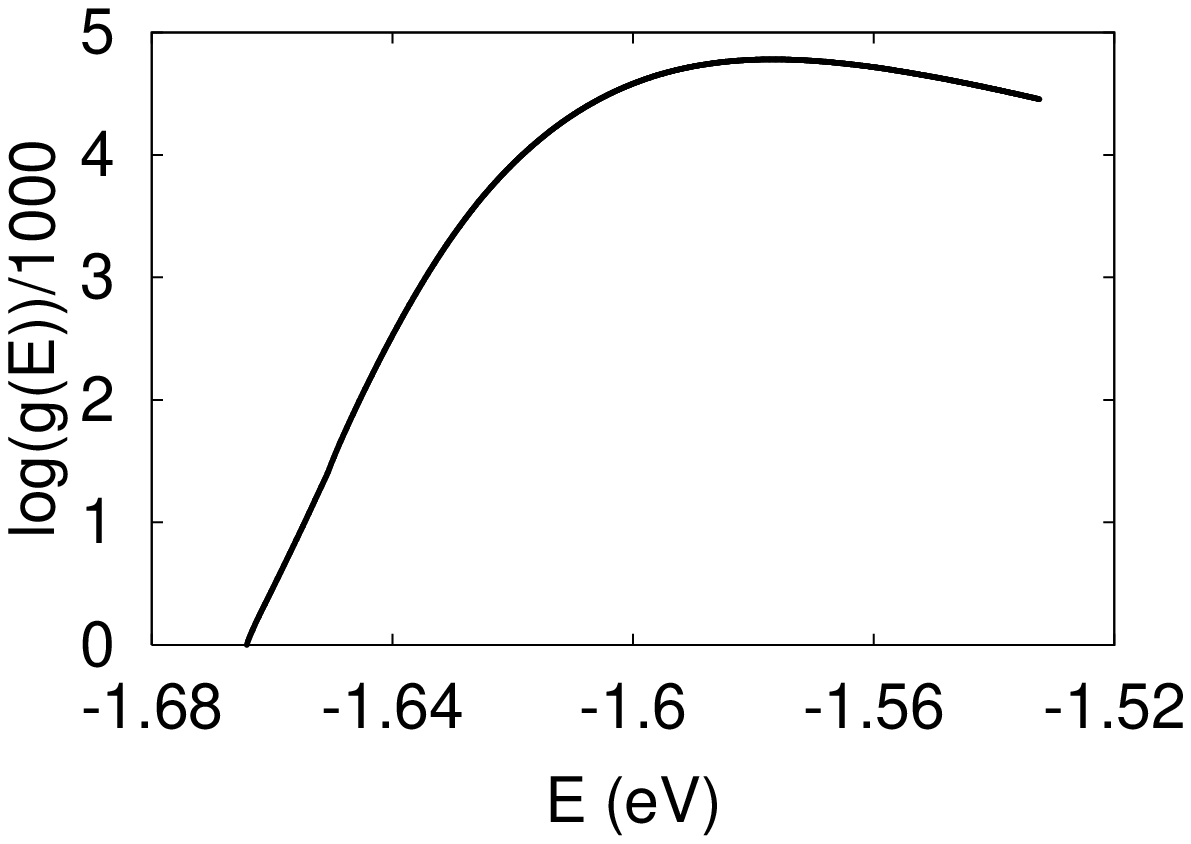}}
\scalebox{0.34}{\includegraphics{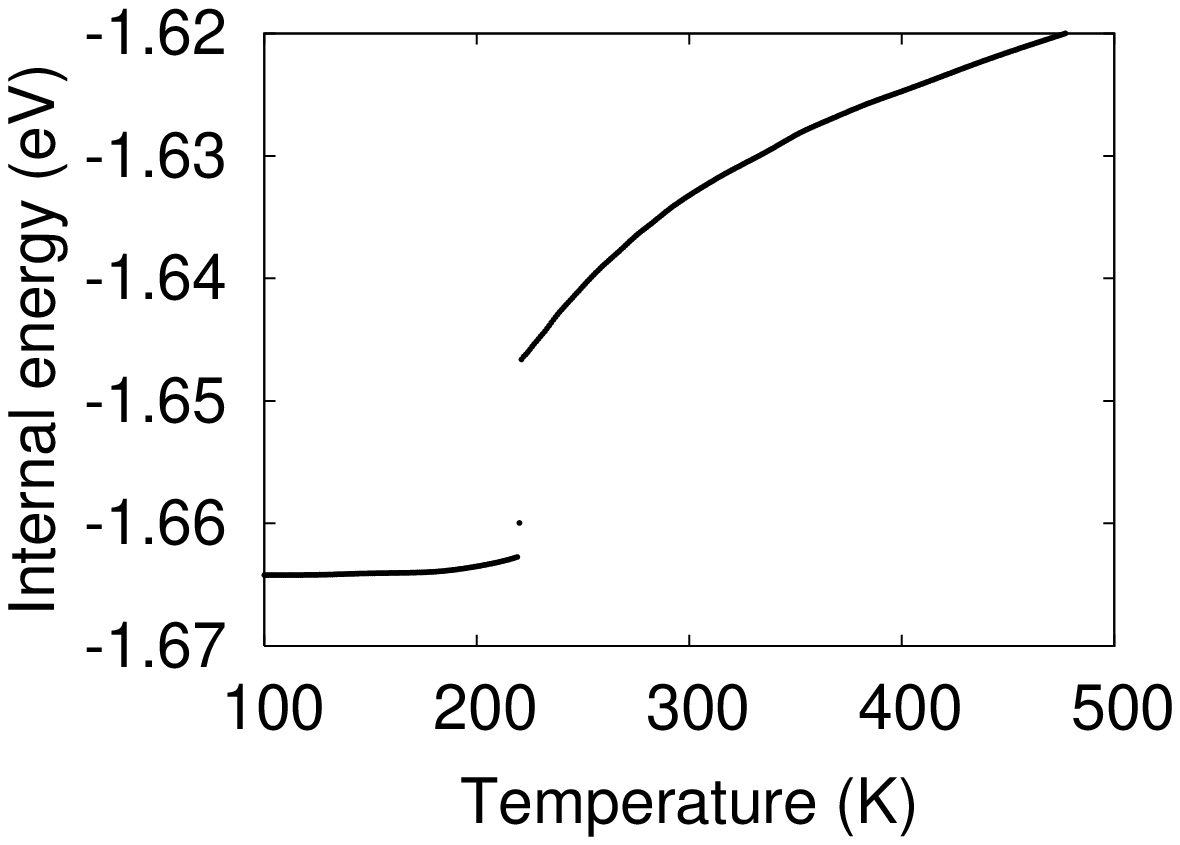}}\scalebox{0.34}{\includegraphics{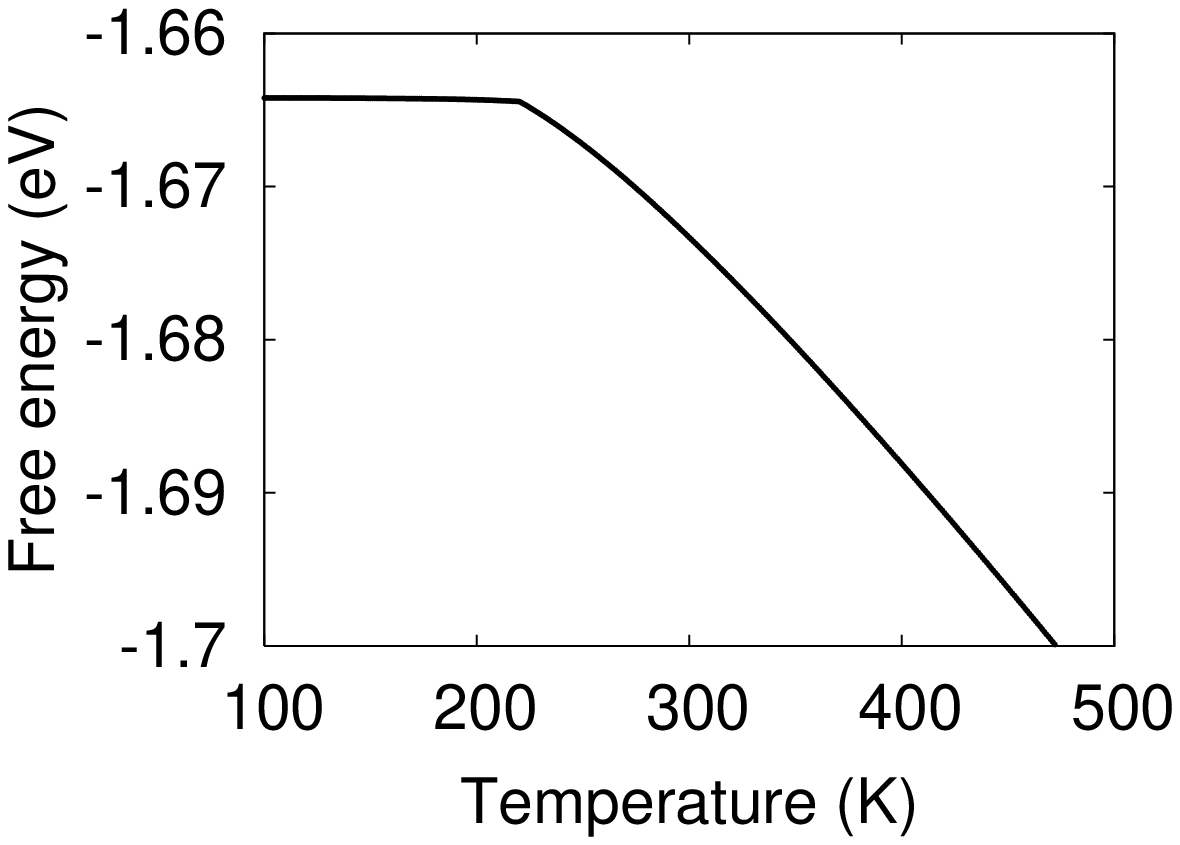}}
\caption{Top panel: Calculated logarithm of the density of configuration states, $g(E)$, for Na on Al(100) at a coverage of 0.2\,monolayers. Lower panels: Internal and free energy of this system as derived from $g(E)$ (from Borg {\em et al.}, 2004).}
\label{fig_naal}
\end{figure}

Using the {\em ab initio} LGH + MC approach the ordered phase and the disorder transition can be successfully reproduced. Pair interactions up to the ninth nearest neighbor and two different trio interactions were included in the LGH expansion. To specifically identify the crucial role played by configurational entropy in the temperature induced order-disorder transition, a specific MC algorithm proposed by Wang and Landau (Wang and Landau, 2001) was employed. In contrast to the above outlined Metropolis algorithm, this scheme affords an explicit calculation of the density of configuration states, $g(E)$, i.e. the number of system configurations with a certain energy $E$. This quantity provides in turn all major thermodynamic functions, e.g., the canonical distribution at a given temperature, $g(E)e^{-E/k_{B}T}$, the free energy, $F(T)=-k_{B}T ln( \sum_{E}g(E)e^{-E/k_{B}T})= -k_{B}T\, ln(Z)$, where $Z$ is the partition function, the internal energy, $U(T)=[\sum_{E}Eg(E)e^{-E/k_{B}T}]/Z$, and the entropy $S=(U-F)/T$. 

Figure~\ref{fig_naal} shows the calculated density of configuration states $g(E)$, together with the internal and free energy derived from it. In the latter two quantities, the abrupt change corresponding to the first-order phase transition obtained at 210\,K can nicely be discerned. In particular, the free energy decreases notably with increasing temperature. The reason for this is clearly the entropic contribution (difference in the free and internal energies), the magnitude of which suddenly increases at the transition temperature and continues to increase steadily thereafter. Taking this configurational entropy into account is therefore (and obviously) the crucial aspect in the simulation and understanding of this order-disorder phase transition, and only the LGH+MC approach with its proper sampling of configuration space can provide it. What the approach does not yield, on the other hand, is {\em how} the phase transition actually takes place microscopically, i.e. how the substitutional Na atoms move their positions by necessarily displacing surface Al atoms, and on what time scale (with what kinetic hindrance) this all happens. For this, one necessarily needs to go beyond a thermodynamic description, and explicitly follow the kinetics of the system over time, which will be the topic of the following section.\\

{\bf First-principles kinetic Monte Carlo simulations} \\

Up to now we had discussed how equilibrium Monte Carlo simulations can be used to explicitly evaluate the partition function, in order to arrive at surface phase diagrams as function of temperature and partial pressures of the surrounding gas-phase. For this, statistical averages over a sequence of appropriately sampled configurations were taken, and it is appealing to also connect some time evolution to this sequence of generated configurations (MC steps). In fact, quite a number of non-equilibrium problems may already be tackled on the basis of this uncalibrated ``MC time'' (Landau and Binder, 2002). The reason why this does not work in general is twofold. First, equilibrium MC is designed to achieve an optimum sampling within configurational phase space. As such, also MC moves that are unphysical like a particle hop from an occupied site to an unoccupied one, hundreds of lattice spacings away may be allowed, if they help to obtain an efficient sampling of the relevant configurations. The remedy for this obstacle is straightforward, though, as one only needs to restrict the possible MC moves to ``physical'' elementary processes. The second reason on the other hand is more involved, as it has to do with the probabilities with which the individual events are executed. In equilibrium MC the forward and backward acceptance probabilities of time-reversed processes like hops back and forth between two sites only have to fulfill the detailed balance criterion, and this is not enough to establish a proper relationship between MC time and ``real time'' (Kang and Weinberg, 1995).

\begin{figure}
\scalebox{0.33}{\includegraphics{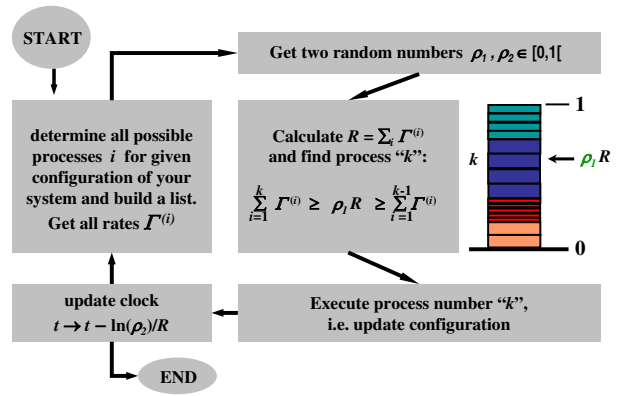}}
\caption{\label{fig_kmcflow}
Flow diagram illustrating the basic steps in a kinetic Monte Carlo simulation. 
i) Loop over all lattice sites of the system and determine the atomic processes that are possible for the current system configuration. Calculate or lookup the corresponding rates. ii) Generate two random numbers, iii) advance the system according to the process selected by the first random number (this could e.g. be moving an atom from one lattice site to a neighboring one, if the corresponding diffusion process was selected). iv) Increment the clock according to the rates and the second random number, as prescribed by an ensemble of Poisson processes,
and v) start all over or stop, if a sufficiently long time span has been simulated.}
\end{figure}

In {\em kinetic Monte Carlo simulations} (kMC) a proper relationship between MC time and real time is achieved by interpreting the Monte Carlo process as providing a numerical solution to the Markovian master equation describing the dynamic system evolution (Bortz, Kalos, and Lebowitz, 1975; Gillespie, 1976; Voter, 1986; Kang and Weinberg, 1989; Fichthorn and Weinberg, 1991). The simulation itself still looks superficially similar to equilibrium Monte Carlo in that a sequence of configurations is generated using random numbers. At each configuration, however, all possible elementary processes and the rates with which they occur are evaluated. Appropriately weighted by these different rates one of the possible processes is then executed randomly to achieve the new system configuration, as sketched in Figure \ref{fig_kmcflow}. This way, the kMC algorithm effectively simulates stochastic processes described by a Poisson distribution, and a direct and unambiguous relationship between kMC time and real time can be established (Fichthorn and Weinberg, 1991). Not only does this open the door to a treatment of the kinetics of non-equilibrium problems. It also does so very efficiently, since the time evolution is actually {\em coarse-grained} to the really decisive rare events, passing over the irrelevant short-time dynamics. Time scales of the order of seconds or longer for mesoscopically-sized systems are therefore readily accessible by kMC simulations (Voter, Montalenti, and Germann, 2002).\\

{\em Insights from MD, MC and kMC} \\

\begin{figure}
\scalebox{0.38}{\includegraphics{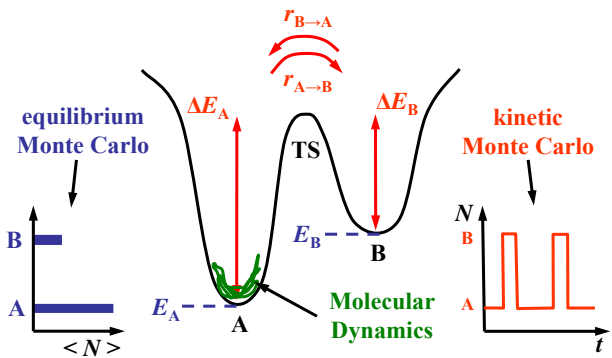}}
\caption{\label{fig_kmcschematic}
Schematic potential energy surface (PES) representing the thermal diffusion of an isolated adsorbate between two stable lattice sites A and B of different stability. A MD simulation would explicitly follow the dynamics of the vibrations around a minimum, and is thus inefficient to address the rare diffusion events happening on a much longer time scale. Equilibrium Monte Carlo simulations provide information about the average thermal occupation of the two sites, $<\!\!\!N\!\!\!>$, based on the depth of the two PES minima ($E_{\rm A}$ and $E_{\rm B}$). Kinetic Monte Carlo simulations follow the ``coarse-grained'' time evolution of the system, $N(t)$, employing the rates for the diffusion events between the minima ($r_{\rm A \rightarrow B}$, $r_{\rm B \rightarrow A}$). For this, PES information not only about the minima, but also about the barrier height at the transition state (TS) between initial and final state is required ($\Delta E_{\rm A}$, $\Delta E_{\rm B}$).}
\end{figure}

To further clarify the different insights provided by molecular dynamics, equilibrium and kinetic Monte Carlo simulations, consider the simple, but typical rare event type model system shown in Figure \ref{fig_kmcschematic}. An isolated adsorbate vibrates at finite temperature $T$ with a frequency on the picosecond time scale and diffuses about every microsecond between two neighboring sites of different stability. In terms of a PES, this situation is described by two stable minima of different depths separated by a sizable barrier. Starting with the particle in any of the two sites, a MD simulation would follow the thermal motion of the adsorbate in detail. In order to do this accurately, timesteps in the femtosecond range are required. Before the first diffusion event can be observed at all, of the order of $10^{9}$ time steps have therefore to be calculated first, in which the particle does nothing but just vibrate around the stable minimum. Computationally this is unfeasible for any but the simplest model systems, and even if it were feasible it would obviously not be an efficient tool to study the long-term time evolution of this system. 

For Monte Carlo simulations on the other hand, the system first has to be mapped onto a lattice. This is unproblematic for the present model and results in two possible system states with the particle being in one or the other minimum. Equilibrium Monte Carlo provides then only time-averaged information about the equilibrated system. For this, a sequence of configurations with the system in either of the two system states is generated, and considering the higher stability of one of the minima, appropriately more configurations with the system in this state are sampled. When taking the average, one arrives at the obvious result that the particle is with a certain higher (Boltzmann-weighted) probability in the lower minimum than in the higher one. 

Real information on the long term time-evolution of the system, i.e. focusing on the rare diffusion events, is finally provided by kinetic Monte Carlo simulations. For this, first the two rates of the diffusion events from one system state to the other and vice versa have to be known. We will describe below that they can be obtained from knowledge of the barrier between the two states and the vibrational properties of the particle in the minima and at the barrier, i.e. from the local curvatures. A lot more information on the PES is therefore required for a kMC simulation than for equilibrium MC, which only needs input about the PES minima. Once the rates are known, a kMC simulation starting from any arbitrary system configuration will first evaluate all possible processes and their rates and then execute one of them with appropriate probability. In the present example this list of events is trivial, since with the particle in either minimum only the diffusion to the other minimum is possible. When the event is executed, on average the time (rate)$^{-1}$ has passed and the clock is advanced accordingly. Note that as described initially, the rare diffusion events happen on a time scale of microseconds, i.e. with only one executed event the system time will be directly incremented by this amount. In other words, the time is {\em coarse-grained} to the rare event time, and all the short-time dynamics (corresponding in the present case to the picosecond vibrations around the minimum) are efficiently contained in the process rate itself. 

Since the barrier seen by the particle when in the shallower minimum is lower than when in the deeper one, cf. Figure \ref{fig_kmcschematic}, the rate to jump into the deeper minimum will correspondingly be higher than the one for the backwards jump. Generating the sequence of configurations, each time more time will therefore have passed after a diffusion event from deep to shallow compared to the reverse process. When taking a long-time average, describing then the equilibrated system, one therefore arrives necessarily at the result that the particle is on average longer in the lower minimum than in the higher one. This is identical to the result provided by equilibrium Monte Carlo, and if only this information is required, the latter technique would most often be the much more efficient way to obtain it. KMC, on the other hand, has the additional advantage of shedding light on   the detailed time-evolution itself, and can in particular also follow the explicit kinetics of systems that are not (or not yet) in thermal equilibrium.

From the discussion of this simple model system, it is clear that the key ingredients of a kMC simulation are the analysis and identification of all possibly relevant elementary processes and the determination of the associated rates. Once this is known, the coarse graining in time achieved in kMC immediately allows to follow the time evolution and the statistical occurrence and interplay of the molecular processes of mesoscopically sized systems up to seconds or longer. As such it is currently the most efficient approach to study long time and larger length scales, while still providing atomistic information. In its original development, kMC was exclusively applied to simplified model systems, employing a few processes with guessed or fitted rates (see e.g. Kang and Weinberg, 1995). The new aspect brought into play by so-called {\em first-principles kMC simulations} (Ruggerone, Ratsch and Scheffler, 1997; Ratsch, Ruggerone and Scheffler, 1998) is that these rates and the processes are directly provided from electronic structure theory calculations, i.e. that the parameters fed into the kMC simulation have a clear microscopic meaning.\\

{\em Getting the processes and their rates} \\

For the rare event type molecular processes mostly encountered in the surface science context, an efficient and reliable way to obtain the individual process rates is transition-state theory (TST) (Glasston, Laidler and Eyring, 1941; 
Vineyard, 1957; Laidler, 1987). The two basic quantities entering this theory are an effective attempt frequency, $\Gamma_{\circ}$, and the minimum energy barrier $\Delta E$ that needs to be overcome for the event to take place, i.e. to bring the system from the initial to the final state. The atomic configuration corresponding to $\Delta E$ is accordingly called the transition state (TS). Within a harmonic approximation, the effective attempt frequency is proportional to the ratio of normal vibrational modes at the initial and transition state. Just like the barrier $\Delta E$, $\Gamma_{\circ}$ is thus also related to properties of the PES, and as such directly amenable to a calculation with electronic structure theory methods like DFT (Ratsch and Scheffler, 1998).

\begin{figure}
\scalebox{0.31}{\includegraphics{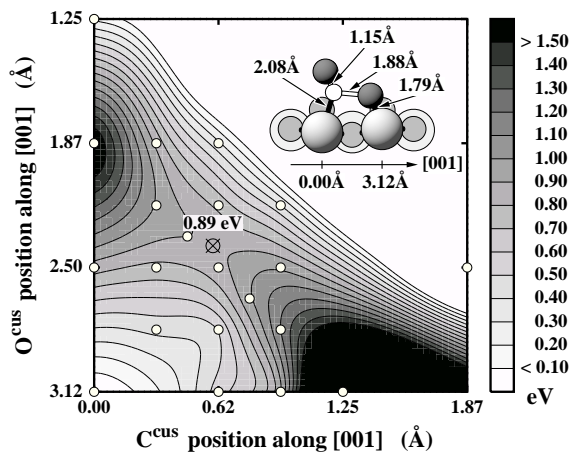}}
\caption{\label{fig_tsgeo}
Calculated DFT-PES of a CO oxidation reaction process at the RuO$_2$(110) model catalyst surface. The high-dimensional PES is projected onto two reaction coordinates, representing two lateral coordinates of the adsorbed O$^{\rm cus}$ and CO$^{\rm cus}$ (cf. Figure \ref{fig_oruphasediag}). The energy zero corresponds to the initial state at (0.00 {\AA}, 3.12 {\AA}), and the transition state is at the saddle point of the PES, yielding a barrier of 0.89\,eV. Details of the corresponding transition state geometry are shown in the inset. Ru = light, large spheres, O = dark, medium spheres, and C = small, white spheres (only the atoms lying in the reaction plane itself are drawn as threedimensional spheres). (From Reuter and Scheffler, 2003b).}
\end{figure}

In the end, the crucial additional PES information required in kMC compared to equilibrium MC is therefore the location of the transition state in form of the PES saddle point along a reaction path of the process. Particularly for high-dimensional PES this is not at all a trivial problem, and the development of efficient and reliable transition-state-search algorithms is a very active area of current research (Henkelman, Johannesson and Jonsson, 2000). For many surface related elementary processes (e.g. diffusion, adsorption, desorption or reaction events) the dimensionality is fortunately not excessive, or can be mapped onto a couple of prominent reaction coordinates as exemplified in Figure \ref{fig_tsgeo}. The identification of the TS and the ensuing calculation of the rate for individual identified elementary processes with TST are then computationally involved, but just feasible. 

\begin{figure}
\scalebox{0.42}{\includegraphics{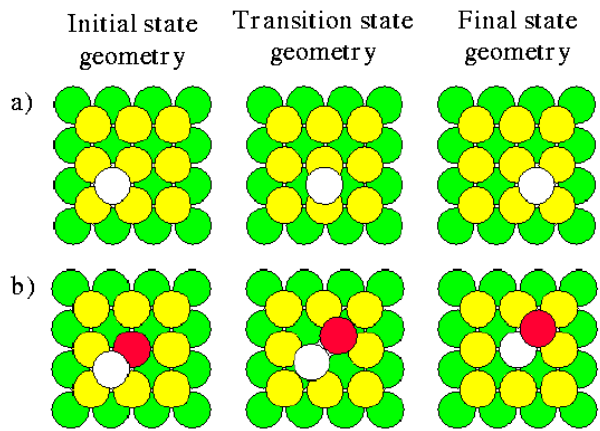}}
\caption{\label{fig_metaldiffusion}
Schematic top view of a fcc(100) surface, explaining diffusion processes of an isolated metal adatom (white circle). a) Diffusion by hopping to a neighboring lattice site, b) diffusion by exchange with a surface atom.}
\end{figure}

This still leaves as a fundamental problem, how the relevant elementary processes for any given system configuration can be identified in the first place. Most TS-search algorithms require not only the automatically provided information of the actual system state, but also knowledge of the final state after the process has taken place (Henkelman, Johannesson and Jonsson, 2000). In other words, quite some insight into the physics of the elementary process is needed in order to determine its rate and include it in the list of possible processes in the kMC simulation. How difficult and non-obvious this can be even for the simplest kind of processes is nicely exemplified by the diffusion of an isolated metal atom over a close-packed surface (Ala-Nissila, Ferrando and Ying, 2002). Such a process is of fundamental importance for the epitaxial growth of metal films, which is a necessary prerequisite in many applications like catalysis, magneto-optic storage media or interconnects in microelectronics. Intuitively, one would expect the surface diffusion to proceed by simple hops from one lattice site to a neighboring lattice site, as illustrated in Figure \ref{fig_metaldiffusion}a for a fcc (100) surface. Having said that, it is in the meanwhile well established that on a number of substrates diffusion does not operate preferentially by such {\em hopping} processes, but by atomic {\em exchange} as explained in Figure \ref{fig_metaldiffusion}b. Here, the adatom replaces a surface atom, and the latter then assumes the adsorption site. Even much more complicated, correlated exchange diffusion processes involving a larger number of surface atoms are currently discussed for some materials. And the complexity increases of course further, when diffusion along island edges, across steps and around defects needs to be treated in detail (Ala-Nissila, Ferrando and Ying, 2002).

While it is therefore straightforward to say that one wants to include e.g. diffusion in a kMC simulation, it can in practice be very involved to identify the individual processes actually contributing to it. Some attempts to automatize the search for the elementary processes possible for a given system configuration are currently undertaken, but in the large majority of first-principles kMC studies performed up to date and in the foreseeable future, the process lists are simply generated by physical insight. This obviously bears the risk of overlooking a potentially relevant molecular process, and on this note this just evolving method has to be seen. Contrary to traditional kMC studies, where an unknown number of real molecular processes is often lumped together into a handful effective processes with optimized rates, first-principles kMC has the advantage, however, that the omission of a relevant elementary process will definitely show up in the simulation results. As such, first experience tells that a much larger number of molecular processes needs to be accounted for in a corresponding modeling ``with microscopic understanding'' compared to traditional empirical kMC (Stampfl {\em et al.}, 2002). In other words, that the statistical interplay determining the observable function of materials takes places between quite a number of different elementary processes, and is therefore often way too complex to be understood by just studying in detail the one or other elementary process alone.\\

{\em Applications to semiconductor growth and catalysis} \\

\begin{figure}
\scalebox{0.55}{\includegraphics{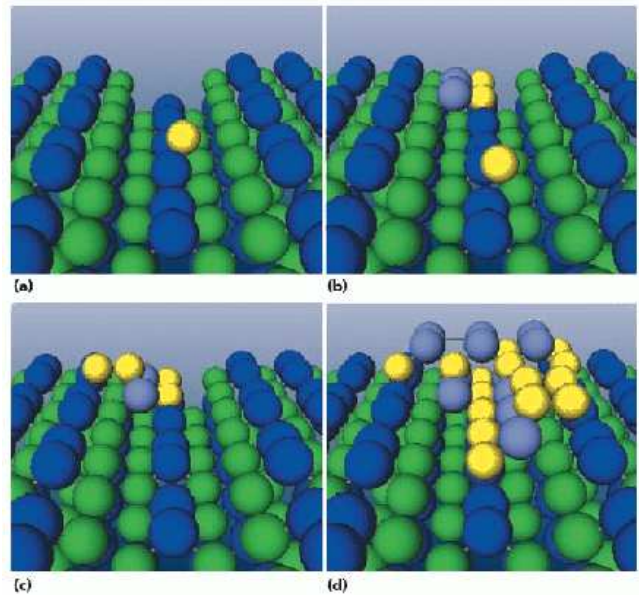}}
\caption{\label{fig_gaaskmc}
Snapshots of characteristic stages during a first-principles kMC simulation of GaAs homoepitaxy. Ga and As substrate atoms appear in green and dark blue, Ga adatoms in yellow, and freshly adsorbed As dimers in light blue. (a) Ga adatoms preferentially wander around in the trenches. (b) Under the growth conditions used here, an As$_2$ molecule adsorbing on a Ga adatom in the trench initiates island formation. (c) Growth proceeds into a new atomic layer via Ga adatoms forming Ga dimers. (d) Eventually, a new layer of arsenic starts to grow, and the island extends itself towards the foreground, while more material attaches along the trench (from Kratzer and Scheffler, 2002).}
\end{figure}

The new quality of and the novel insights that can be gained by mesoscopic first-principles kMC simulations was first demonstrated in the area of nucleation and growth in metal and semiconductor epitaxy (Ruggerone, Ratsch and Scheffler, 1997; Ratsch, Ruggerone and Scheffler, 1998; Ovesson, Bogicevic and Lundqvist, 1999; Fichthorn and Scheffler, 2000; Kratzer and Scheffler, 2001; Kratzer and Scheffler, 2002; Kratzer, Penev and Scheffler, 2002). As one example from this field we return to the GaAs(001) surface already discussed in the context of the chemical potential plots. As apparent from Figure \ref{fig_sunghoon}, the so-called $\beta2(2\times 4)$ reconstruction represents the most stable phase under moderately As-rich conditions, which are typically employed in the molecular beam epitaxy (MBE) growth of this material. Aiming at an atomic-scale understanding of this technologically most relevant process, first-principles LGH + kMC simulations were performed, including the deposition of As$_2$ and Ga from the gas-phase, as well as diffusion on this complex $\beta2(2\times 4)$ semiconductor surface. In order to reach a trustworthy modeling, the consideration of more than 30 different elementary processes was found to be necessary, underlining our general message that complex material's properties cannot be understood by analyzing isolated molecular processes alone. Snapshots of characteristic stages during a typical simulation at realistic deposition fluxes and temperature are given in Figure \ref{fig_gaaskmc}. They show a small part of the total mesoscopic simulation area, focusing on one ``trench'' of the $\beta2(2\times 4)$ reconstruction. At the chosen conditions, island nucleation is observed in these reconstructed surface trenches, which is followed by growth along the trench, thereby extending into a new layer.

\begin{figure}
\scalebox{0.57}{\includegraphics{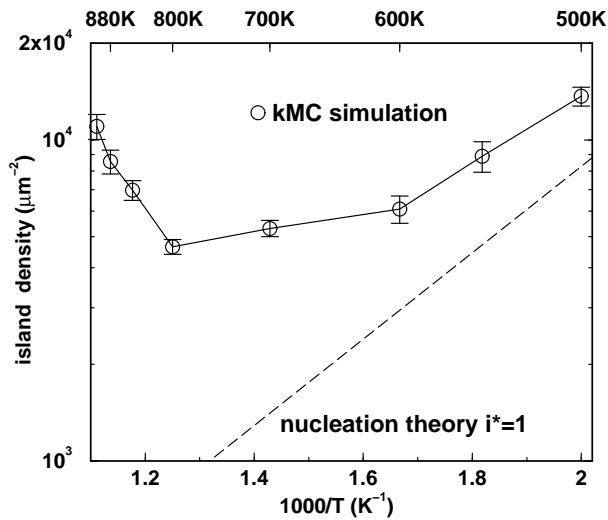}}
\caption{\label{fig_gaasislands}
Saturation island density corresponding to steady-state MBE of GaAs as a function of the inverse growth temperature. The dashed line shows the prediction of classical nucleation theory (CNT) for diffusion-limited attachment and a critical nucleus size equal to 1. The significant deviation at higher temperatures is caused by arsenic losses due to desorption, which is not considered in CNT (from Kratzer, Penev and Scheffler, 2002).}
\end{figure}

Monitoring the density of the nucleated islands in huge simulation cells ($160 \times 320$ surface lattice constants), a saturation indicating the beginning of steady-state growth is only reached after simulation times of the order of seconds for quite a range of temperatures. Obviously, neither such system sizes, nor time scales would have been accessible by direct electronic structure theory calculations combined e.g. with MD simulations. In the ensuing steady-state growth, attachment of a deposited Ga atom to an existing island typically takes place before the adatom could take part in a new nucleation event. This leads to a very small nucleation rate that is counterbalanced by a simultaneous decrease in the number of islands due to coalescence. The resulting constant island density during steady-state growth is plotted in Figure \ref{fig_gaasislands} for a range of technologically relevant temperatures. At the lower end around 500-600\,K, this density decreases, as is consistent with the frequently employed classical nucleation theory (CNT). Under these conditions, the island morphology is predominantly determined by Ga surface diffusion alone, i.e. it may be understood on the basis of one molecular process class. Around 600\,K the island density becomes almost constant, however, and even increases again above around 800\,K. The determined magnitude is then orders of magnitude away from the prediction of CNT, cf. Figure \ref{fig_gaasislands}, but in very good agreement with existing experimental data. The reason for this unusual behavior is that the adsorption of As$_2$ molecules at reactive surface sites becomes reversible at these elevated temperatures. The initially formed Ga-As-As-Ga$_2$ complexes required for nucleation, cf. Figure \ref{fig_gaaskmc}b, become unstable against As$_2$ desorption, and a decreasing fraction of them can stabilize into larger aggregates. Due to the contribution of the decaying complexes, an effectively higher density of mobile Ga adatoms results at the surface, which in turn yields a higher nucleation rate of new islands. The temperature window around 700-800\,K, which is frequently used by MBE crystal growers, may therefore be understood as permitting a compromise between high Ga adatom mobility and stability of As complexes that leads to a low island density and correspondingly smooth films. 

Exactly under the technologically most relevant conditions, the desired functionality of the surface results therefore from the concerted interdependence of distinct molecular processes, i.e. in this case diffusion, adsorption and desorption. To further show that this is to our opinion more the rule than an exception in materials science applications, we return in the remainder of this section to the field of heterogeneous catalysis. Here, the conversion of reactants into products by means of surface chemical reactions ($A + B \rightarrow C$) adds another qualitatively different class of processes to the statistical interplay. In the context of the thermodynamic chemical potential plots we had already discussed that these on-going catalytic reactions at the surface continuously consume the adsorbed reactants, driving the surface populations away from their equilibrium value. If this has a significant effect, presumably e.g. in regions of very high catalytic activity, the average surface coverage and structure does even under steady-state operation never reach its equilibrium with the surrounding reactant gas-phase, and must thence be modeled by explicitly accounting for the surface kinetics (Hansen and Neurock, 1999; Hansen and Neurock, 2000; Reuter, Frenkel and Scheffler, 2004). 

\begin{figure}
\scalebox{0.6}{\includegraphics{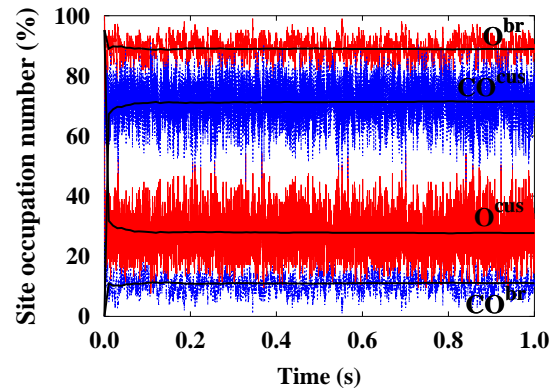}}
\caption{\label{fig_oruoccupation}
Time evolution of the site occupation by O and CO of the two prominent adsorption sites of the RuO$_2$(110) model catalyst surface shown in Figure \ref{fig_oruphasediag}. The temperature and pressure conditions chosen ($T=600$\,K, $p_{\rm CO} = 20$\,atm, $p_{\rm O_2} = 1$\,atm) correspond to an optimum catalytic performance. Under these conditions kinetics builds up a steady-state surface population in which O and CO compete for either site type at the surface, as reflected by the strong fluctuations in the site occupations. Note the extended time scale, also for the ``induction period'' until the steady-state populations are reached when starting from a purely oxygen covered surface (from Reuter, Frenkel and Scheffler, 2004).}
\end{figure}

\begin{figure}
\scalebox{0.46}{\includegraphics{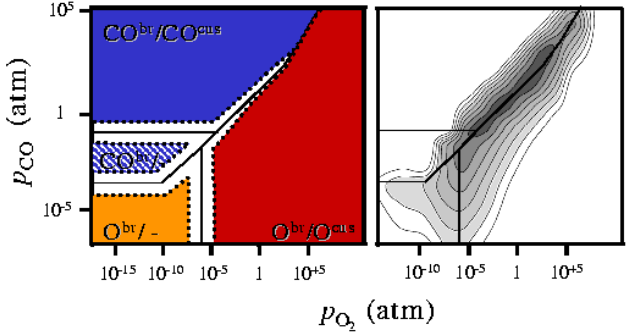}}
\caption{\label{fig_orukMC}
Left panel: Steady state surface structures of RuO$_2$(110) in an O$_2$/CO environment obtained by first-principles kMC calculations at $T = 600$\,K. In all non-white areas, the average site occupation is dominated ($> 90$\,\%) by one species, and the site nomenclature is the same as in Figure \ref{fig_oruphasediag}, where the same surface structure was addressed within the {\em ab initio} atomistic thermodynamics approach. Right panel: Map of the corresponding catalytic CO oxidation activity measured as so-called turn-over frequencies (TOFs), i.e. CO$_2$ conversion per ${\rm cm}^{2}$ and second: White areas have a TOF\,$< 10^{11} {\rm cm}^{-2}{\rm s}^{-1}$, and each increasing gray level represents one order of magnitude higher activity. The highest catalytic activity (black region, TOF $> 10^{17}\,{\rm cm}^{-2}{\rm s}^{-1}$) is narrowly concentrated around the phase coexistence region that was already suggested by the thermodynamic treatment (from Reuter, Frenkel and Scheffler, 2004).}
\end{figure}

In terms of kMC, this means that in addition to the diffusion, adsorption and desorption of the reactants and products, also reaction events have to be considered. For the case of CO oxidation as one of the central reactions taking place in our car catalytic converters, this translates into the conversion of adsorbed O and CO into CO$_2$. Even for the afore discussed, moderately complex model catalyst RuO$_2$(110), again close to 30 elementary processes result, comprising both adsorption to and desorption from the two prominent site-types at the surface (br and cus, cf. Figure \ref{fig_oruphasediag}), as well as diffusion between any nearest neighbor site-combination (br$\rightarrow$br, br$\rightarrow$cus, cus$\rightarrow$br, cus$\rightarrow$cus). Finally, reaction events account for the catalytic activity and are possible whenever O and CO are similarly adsorbed in any nearest neighbor site-combination. For given temperature and reactant pressures, the corresponding kMC simulations are then first run until steady-state conditions are reached, and the average surface populations are thereafter evaluated over sufficiently long times. We note that even for elevated temperatures, both time periods may again largely exceed the time span accessible by current MD techniques as exemplified in Figure \ref{fig_oruoccupation}. The obtained steady-state average surface populations at $T=600$\,K are shown in Figure \ref{fig_orukMC} as a function of the gas-phase partial pressures. Comparing with the surface phase diagram of Figure \ref{fig_oruphasediag} from {\em ab initio} atomistic thermodynamics, i.e. neglecting the effect of the on-going catalytic reactions at the surface, similarities, but also the expected significant differences under some environmental conditions can be discerned. 

The differences affect most prominently the presence of oxygen at the br sites, where it is much more strongly bound than CO. For the thermodynamic approach only the ratio of adsorption to desorption matters, and due to the ensuing very low desorption rate, O$^{\rm br}$ is correspondingly stabilized even when there is much more CO in the gas-phase than O$_2$ (left upper part of Figure \ref{fig_oruphasediag}). The surface reactions, on the other hand, provide a very efficient means of removing this O$^{\rm br}$ species that is not accounted for in the thermodynamic treatment. As net result, under most CO-rich conditions in the gas-phase, oxygen is faster consumed by the reaction than it can be replenished from the gas-phase. The kMC simulations covering this effect yield then a much lower surface concentration of O$^{\rm br}$, and in turn show a much larger stability range of surface structures with CO$^{\rm br}$ at the surface (blue and hatched blue regions). It is particularly interesting to notice, that this yields a stability region of a surface structure consisting of only adsorbed CO at br sites that does not exist in the thermodynamic phase diagram at all, cf. Figure \ref{fig_oruphasediag}. The corresponding CO$^{\rm br}$/- ``phase'' (hatched blue region) is thus a stable structure with defined average surface population that is entirely stabilized by the kinetics of this open catalytic system.

These differences were conceptually anticipated in the thermodynamic phase diagram, and qualitatively delineated by the hatched regions in Figure \ref{fig_oruphasediag}. Due to the vicinity to a phase transition and the ensuing enhanced dynamics at the surface, these regions were also considered as potential candidates for highly efficient catalytic activity. This is in fact confirmed by the first-principles kMC simulations as shown in the right panel of Figure \ref{fig_orukMC}. Since the detailed statistics of all elementary processes is explicitly accounted for in the latter type simulations, it is straightforward to also evaluate the average occurrence of the reaction events over long time periods as a measure of the catalytic activity. The obtained so-called turnover frequencies (TOF, in units of formed CO$_2$ per cm$^{2}$ per second) are indeed narrowly peaked around the phase coexistence line, where the kinetics builds up a surface population in which O and CO compete for either site type at the surface. This competition is in fact nicely reflected by the large fluctuations in the surface populations apparent in Figure \ref{fig_oruoccupation}. The partial pressures and temperatures corresponding to this high activity ``phase'', and even the absolute TOF values under these conditions, agree extremely well with detailed experimental studies measuring the steady-state activity in the temperature range from 300-600\,K and both at high pressures and in UHV. Interestingly, under the conditions of highest catalytic performance it is not the reaction with the highest rate (lowest barrier) that dominates the activity. Although the particular elementary process itself exhibits very suitable properties for catalysis, it occurs too rarely in the full concert of all possible events to decisively affect the observable macroscopic functionality. This emphasizes again the importance of the statistical interplay and the novel level of understanding that can only be provided by first-principles based mesoscopic studies.\\

{\bf Outlook} \\

As highlighted by the few examples from surface physics discussed above, many materials` properties and functions arise out of the interplay of a large number of distinct molecular processes. Theoretical approaches aiming at an atomic-scale understanding and predictive modeling of such phenomena have therefore to achieve both an accurate description of the individual elementary processes at the electronic regime {\em and} a proper treatment of how they act together on the mesoscopic level. We have sketched the current status and future direction of some emerging methods which correspondingly try to combine electronic structure theory with concepts from statistical mechanics and thermodynamics. The results already achieved with these techniques give a clear indication of the new quality and novelty of insights that can be gained by such descriptions. On the other hand, it is also apparent that we are only at the beginning of a successful bridging of the micro- to mesoscopic transition in the multi-scale materials modeling endeavor. Some of the major conceptual challenges we see at present that need to be tackled when applying these schemes to more complex systems have been touched in this chapter. They may be summarized under the keywords {\em accuracy, mapping and efficiency}, and as outlook we briefly comment further on them.

{\em Accuracy}: The reliability of the statistical treatment depends predominantly on the accuracy of the description of the individual molecular processes that are input to it. For the mesoscopic methods themselves it makes in fact no difference, whether the underlying PES comes from a semi-empirical potential or from first-principles calculations, but the predictive power of the obtained results (and the physical meaning of the parameters) will obviously be significantly different. In this respect we only mention two somehow diverging aspects. For the interplay of several (possibly competing) molecular processes, an accurate {\em absolute} description of each individual process e.g. in form of a rate for kMC simulations may be less important than the {\em relative} ordering among the processes as e.g. provided by the correct trend in their energetics. In this case, the frequently requested {\em chemical accuracy} in the description of single processes could be a misleading concept, and modest errors in the PES would tend to cancel (or compensate each other) in the statistical mechanics part. For the particular case of DFT as the current workhorse of electronic structure theories this could mean that the present uncertainties due to the approximate treatment of electronic exchange and correlation are less problematic than hitherto often assumed. On the other hand, in other cases where for example one process strongly dominates the concerted interplay, such a cancellation will certainly not occur. Then, a more accurate description of this process will be required than can be provided by the exchange-correlation functionals in DFT that are available today. Improved descriptions based on wave-function methods and on local corrections to DFT exist or are being developed, but come so far at a high computational cost. Assessing what kind of accuracy is required for which process under which system state, possibly achieved by evolutionary schemes based on gradually improving PES descriptions, will therefore play a central role in making atomistic statistical mechanics methods computationally feasible for increasingly complex systems.

{\em Mapping:} The configuration space of most materials science problems is exceedingly large. In order to arrive at meaningful statistics, even the most efficient sampling of such spaces still requires (at present and in the foreseeable future) a number of PES evaluations that is prohibitively large to be directly provided by first-principles calculations. This problem is mostly circumvented by mapping the actual system onto a coarse-grained lattice model, in which the real Hamiltonian is approximated by discretized expansions e.g. in certain interactions (LGH) or elementary processes (kMC). The expansions are then first parametrized by the first-principles calculations, while the statistical mechanics problem is thereafter solved exploiting the fast evaluations of the model Hamiltonians. Since in practice these expansions can only comprise a finite number of terms, the mapping procedure intrinsically bears the problem of overlooking a relevant interaction or process. Such an omission can obviously jeopardize the validity of the complete statistical simulation, and there are at present no fool-proof or practical, let alone automatized schemes as to which terms to include in the expansion, neither how to judge on the convergence of the latter. In particular when going to more complex systems the present ``hand-made'' expansions that are mostly based on educated guesses will become increasingly cumbersome. Eventually, the complexity of the system may become so large, that even the mapping onto a discretized lattice itself will be problematic. Overcoming these limitations may be achieved by adaptive, self-refining approaches, and will certainly be of paramount importance to ensure the general applicability of the atomistic statistical techniques.

{\em Efficiency:} Even if an accurate mapping onto a model Hamiltonian is achieved, the sampling of the huge configuration spaces will still put increasing demands on the statistical mechanics treatment. In the examples discussed above, the actual evaluation of the system partition function e.g. by (k)MC simulations is a small add-on compared to the computational cost of the underlying DFT calculations. With increasing system complexity, different problems and an increasing number of processes this may change eventually, requiring the use of more efficient sampling schemes. A major challenge for increasing efficiency is for example the treatment of processes operating at largely different time scales. The computational cost of a certain time span in kMC simulations is dictated by the fastest process in the system, while the slowest process governs what total time period needs actually to be covered. If both process scales differ largely, kMC becomes expensive. Remedy may e.g. be provided by assuming the fast process to be always equilibrated at the time scale of the slow one, and correspondingly an appropriate mixing of equilibrium MC with kMC simulations may significantly increase the efficiency (as already done in nowadays TPD simulations). Alternatively, the fast process could not be explicitly considered anymore on the atomistic level, and only its effect incorporated into the remaining processes. 

Obviously, with such a grouping of processes one approaches already the meso- to macroscopic transition, gradually giving up the atomistic description in favor of a more coarse-grained or even continuum modeling. The crucial point to note here is that such a transition is done in a controlled and hierarchical manner, i.e. necessarily as the outcome and understanding from the analysis of the statistical interplay at the mesoscopic level. This is therefore in marked contrast to e.g. the frequently employed rate equation approach in heterogeneous catalysis modeling, where macroscopic differential equations are directly fed with microscopic parameters. If the latter are simply fitted to reproduce some experimental data, at best a qualitative description can be achieved anyway. If really microscopically meaningful parameters are to be used, one does not know which of the many in principle possible elementary processes to consider. Simple-minded ``intuitive'' approaches like e.g. parametrizing the reaction equation with the data from the reaction process with the highest rate may be questionable in view of the results described above. This process may never occur in the full concert of the other processes, or it may only contribute under particular environmental conditions, or be significantly enhanced or suppressed due to an intricate interplay with another process. All this can only be filtered out by the statistical mechanics at the mesoscopic level, and can therefore not be grasped by the traditional rate equation approach omitting this intermediate time and length scale regime. 

The two key features of the atomistic statistical schemes reviewed here are in summary that they treat the statistical interplay of the possible molecular processes, and that these processes have a well-defined microscopic meaning, i.e. they are described by parameters that are provided by first-principles calculations. This distinguishes these techniques from approaches where molecular process parameters are either directly put into macroscopic equations neglecting the interplay, or where only effective processes with optimized parameters are employed in the statistical simulations. In the latter case, the individual processes loose their well-defined microscopic meaning and typically represent an unspecified lump sum of not further resolved processes. Both the clear cut microscopic meaning of the individual processes and their interplay are, however, decisive for the transferability and predictive nature of the obtained results. Furthermore, it is also precisely these two ingredients that ensure the possibility of reverse-mapping, i.e. the unambiguous tracing back of the microscopic origin of (appealing) materials` properties identified at the meso- or macroscopic modeling level. We are convinced that primarily the latter point will be crucial when trying to overcome the present trial and error based system engineering in materials sciences in the near future. An advancement based on {\em understanding} requires theories that straddle various traditional disciplines. The approaches discussed here employ methods from various areas of electronic structure theory (physics as well as chemistry), statistical mechanics, mathematics, materials science, and computer science. This high interdisciplinarity makes the field challenging, but is also part of the reason why it is exciting, timely and full with future perspectives.

\vspace{0.5cm}
{\bf Bibliography} \\

\noindent Adams, D.L. 1996.
New phenomena in the adsorption of alkali metals on Al surfaces,
{\em Appl. Phys. A 62, 123}.

\noindent Ala-Nissila, T., Ferrando, R. and Ying, S.C. 2002.
Collective and single particle diffusion on surfaces,
{\em Adv. in Physics 51, 949}.

\noindent Allen, M.P. and Tildesley, D.J. 1997. 
{\em Computer Simulation of Liquids},
Oxford University Press, Oxford.

\noindent Borg, M., Stampfl, C., Mikkelsen, A., Gustafson, J.,
Lundgren, E., Scheffler, M., and Andersen, J.N. 2004.
Phase diagram of Al-Na surface alloys from first principles, 
{\em to be published}.

\noindent Bortz, A.B., Kalos, M.H. and Lebowitz, J.L. 1975.
New algorithm for Monte Carlo simulation of ising spin systems,
{\em J. Comp. Phys. 17, 10}.

\noindent Car, R. and Parrinello, M. 1985.
Unified approach for molecular dynamics and density-functional theory,
{\em Phys. Rev. Lett. 55, 2471}.

\noindent Darling, G.R. and Holloway, S. 1995.
The dissociation of diatomic molecules at surfaces,
{\em Rep. Prog. Phys. 58, 1595-1672}.

\noindent De Fontaine, D. 1994. 
In: Turchi, P.E.A. and Gonis, A (eds)
{\em Statics and dynamics of alloy phase transformations},
NATO ASI Series, Plenum Press, New York.

\noindent Diehl, R.D. and Mc Grath, R. 1997.
Current progress in understanding alkali metal adsorption on metal surfaces,
{\em J. Phys. Cond. Mat. 9, 951}.

\noindent Dreizler, R.M. and Gross, E.K.U. 1990.
{\em Density Functional Theory}, 
Springer, Berlin.

\noindent Duke, C.B. 1996.
Semiconductor surface reconstruction: The structural chemistry of 
two-dimensional surface compounds.
{\em Chem. Rev. 96, 1237}.

\noindent Engel, T. and Ertl, G. 1982.
Oxidation of carbon monoxide. In: King, D.A. and Woodruff, D.P. (eds)
{\em The chemical physics of solid surfaces and heterogeneous catalysis},
Elsevier, Amsterdam.

\noindent Ertl, G., Kn\"ozinger, H. and Weitkamp J. (eds) 1997.
{\em Handbook of heterogeneous catalysis}, 
Wiley, New York.

\noindent Ertl, G. 2002.
Heterogeneous catalysis on the atomic scale,
{\em J. Mol. Catal. A 182, 5}.

\noindent Fichthorn, K.A. and  Weinberg, W.H. 1991.
Theoretical foundations of dynamical Monte Carlo simulations,
{\em J. Chem. Phys. 95, 1090}.

\noindent Fichthorn, K.A. and Scheffler, M. 2000.
Island nucleation in thin-film epitaxy: A first-principles investigation,
{\em Phys. Rev. Lett. 84, 5371}.

\noindent Frenkel, D. and Smit, B. 2002.
{\em Understanding Molecular Simulation}, 2nd edn.,
Academic Press, San Diego.

\noindent Galli, G. and Pasquarello, A. 1993.
First-principle molecular dynamics. In: Allen, M.P. and Tildesley, D.J. (eds.)
{\em Computer simulations in chemical physics}, 
Kluwer, Dordrecht.

\noindent Gillespie, D.T. 1976.
General method for numerically simulating stochastic time evolution of
coupled chemical reactions,
{\em J. Comp. Phys. 22, 403}.

\noindent Glasston, S., Laidler, K.J. and Eyring, H. 1941.
{\em The Theory of Rate Processes}, 
McGraw-Hill, New York.

\noindent Gross, A. 1998.
Reactions at surfaces studied by ab initio dynamics calculations,
{\em Surf. Sci. Rep. 32, 293}.

\noindent Hansen, E.W. and Neurock, M. 1999.
Modeling surface kinetics with first-principles-based molecular simulation,
{\em Chem. Eng. Sci. 54, 3411}.

\noindent Hansen, E.W. and Neurock, M. 2000.
First-principles-based Monte Carlo simulation of ethylene hydrogenation kinetics on Pd,
{\em J. Catal. 196, 241}. 

\noindent Hendriksen, B.L.M., Bobaru, S.C. and Frenken, J.W.M. 2003.
Oscillatory CO oxidation on Pd(100) studied with in situ scanning
tunnelling microscopy,
{\em Surf. Sci. 552, 229}.

\noindent Henkelman, G., Johannesson, G. and Jonsson, H. 2000.
Methods for finding saddle points and minimum energy paths. In: Schwartz, S.D. (ed.) {\em Progress on theoretical chemistry and physics}, 
Kluwer, New York.

\noindent Hohenberg, P. and Kohn, W. 1964.  
Inhomogeneous electron gas,
{\em Phys. Rev. B 136, 864}.

\noindent Kang, H.C. and Weinberg, W.H. 1989.
Dynamic Monte Carlo with a proper energy barrier: Surface diffusion and two-dimensional domain ordering,
{\em J. Chem. Phys. 90, 2824}.

\noindent Kang, H.C. and Weinberg, W.H. 1995.
Modeling the kinetics of heterogeneous catalysis,
{\em Chem. Rev. 95, 667}.

\noindent Kaxiras, E., Bar-Yam, Y., Joannopoulos, J.D., and Pandey, K.C. 1987.
Ab initio theory of polar semiconductor surfaces. I. Methodology and
the (22) reconstructions of GaAs(111),
{\em Phys. Rev. B 35, 9625}.

\noindent Koh, S.-J. and Ehrlich, G. 1999.
Pair- and many-atom interactions in the cohesion of surface clusters: Pd$_{x}$
and Ir$_{x}$ on W(110),
{\em Phys. Rev. B 60, 5981}.

\noindent Kohn, W. and Sham, L. 1965.
Self consistent equations including exchange and correlation effects,
{\em Phys. Rev. A 140, 1133}.

\noindent Kratzer, P. and Scheffler, M. 2001.
Surface knowledge: Toward a predictive theory of materials,
{\em Comp. in Science and Engineering 3 (6), 16}.

\noindent Kratzer, P. and Scheffler, M. 2002.
Reaction-limited island nucleation in molecular beam epitaxy of compound semiconductors.
{\em Phys. Rev. Lett. 88, 036102}.

\noindent Kratzer, P., Penev, E., and Scheffler, M. 2002.
First-principles studies of kinetics in epitaxial growth of III-V semiconductors.
{\em Appl. Phys. A 75, 79}.

\noindent Kreuzer, H.J. and Payne, S.H. 2000.
Theoretical approaches to the kinetics of adsorption, desorption and reactions at surfaces. In: Borowko, M. (ed.) {\em Computational methods in surface and colloid}, 
Marcel Dekker, New York.

\noindent Kroes, G.J. 1999.
Six-dimensional quantum dynamics of dissociative chemisorption of H$_2$ on
Metal surfaces.
{\em Prog. Surf. Sci. 60, 1}.

\noindent Laidler, K.J. 1987.
{\em Chemical kinetics}, 
Harper and Row, New York.

\noindent Landau, D.P. and Binder, K. 2002.
{\em A guide to Monte Carlo simulations in statistical physics},
Cambridge University Press, Cambridge.

\noindent Lee, S.-H., Moritz, W., and Scheffler, M. 2000.

GaAs(001) under conditions of low As pressure: Edvidence for a novel surface geometry, 
{\em Phys. Rev. Lett. 85, 3890}.

\noindent Li, W.-X., Stampfl, C., and Scheffler, M. 2003a.
Insights into the function of silver as an oxidation catalyst by ab initio atomistic thermodynamics,
{\em Phys. Rev. B 68, 16541}. 

\noindent Li, W.-X., Stampfl, C. and Scheffler, M. 2003b.
Why is a noble metal catalytically active? The role of the O-Ag interaction in the function of silver as an oxidation catalyst,
{\em Phys. Rev. Lett. 90, 256102}.

\noindent {\L}odzianan, Z. and N{\o}rskov, J.K. 2003.
Stability of the hydroxylated (0001) surface of Al$_{2}$O$_{3}$,
{\em J. Chem. Phys. 118, 11179}.

\noindent Lundgren, E., Kresse, G., Klein, C., Borg, M.,
Andersen, J.N., De Santis, M., Gauthier, Y., Konvicka, C.,
Schmid, M., and Varga, P. 2002.
Two-dimensional oxide on Pd(111),
{\em Phys. Rev. Lett. 88, 246103}.

\noindent Lundgren, E., Gustafson, J., Mikkelsen, A., Andersen, J.N.,
Stierle, A., Dosch, H., Todorova, M., Rogal, J., Reuter, K., and Scheffler, M. 2004.
Kinetic hindrance during the initial oxidation of Pd(100) at ambient pressures,
{\em Phys. Rev. Lett. 92, 046101}.

\noindent Masel, R.I. 1996.
{\em Principles of adsorption and reaction on solid surfaces}, 
Wiley, New York.

\noindent McEwen, J.-S., Payne, S.H., and Stampfl, C. 2002.
Phase diagram of O/Ru(0001) from first principles,
{\em Chem. Phys. Lett. 361, 317}.

\noindent Mc Quarrie, D.A. 1976.
{\em Statistical Mechanics}, 
Harper and Row, New York. 

\noindent Metropolis, N., Rosenbluth, A.W., Rosenbluth, M.N., Teller, A.H., and Teller, E. 1953.
Equation of state calculations by fast computing machines,
{\em J. Chem. Phys. 21, 1087}.

\noindent Michaelides, A., Bocquet, M.L., Sautet, P., Alavi, A., and King, D.A. 2003.
Structures and thermodynamic phase transitions for oxygen and silver oxide
phases on Ag\{111\},
{\em Chem. Phys. Lett. 367, 344}.

\noindent \"{O}sterlund, L., Pedersen, M.{\O}., Stensgaard, I., L\ae gsgaard, E., and Besenbacher, F. 1999. 
Quantitative determination of adsorbate-adsorbate interactions,
{\em Phys. Rev. Lett. 83, 4812}.

\noindent Over, H. and Muhler, M. 2003.
Catalytic CO oxidation over ruthenium -- bridging the pressure gap,
{\em Prog. Surf. Sci. 72, 3}.

\noindent Ovesson, S., Bogicevic, A., and Lundqvist, B.I. 1999.
Origin of compact triangular islands in metal-on-metal growth,
{\em Phys. Rev. Lett. 83, 2608}.

\noindent Parr, R.G. and Yang, W. 1989.
{\em Density functional theory of atoms and molecules},
Oxford University Press, New York.

\noindent Payne, M.C., Teter, M.P., Allan, D.C., Arias, T.A., and Joannopoulos, J.D. 1992.
Iterative minimization techniques for {\em ab initio} total energy calculations: Molecular dynamics and conjugate gradients,
{\em Rev. Mod. Phys. 64, 1045}.

\noindent Payne, S.H., Kreuzer, H.J., Frie, W., Hammer, L., and Heinz, K. 1999.
Adsorption and desorption of hydrogen on Rh(311) and comparison with other Rh surfaces, 
{\em Surf. Sci. 421, 279}.

\noindent Piercy, P., De'Bell, K., and Pfn\"{u}r, H. 1992.
Phase diagram and critical behavior of the adsorption system O/Ru(001):
Comparison with lattice-gas models, 
{\em Phys. Rev. B 45, 1869}.

\noindent Qian, G.-X., Martin, R.M., and Chadi, D.J. 1988.
First-principles study of the atomic reconstructions and energies of
Ga- and As-stabilized GaAs(100) surfaces,
{\em Phys. Rev. B 38, 7649}.

\noindent Ratsch, C., Ruggerone, P. and Scheffler, M. 1998.
Study of strain and temperature dependence of metal epitaxy.
In: Zhang, Z. and Lagally, M.G. (eds.) {\em Morphological organization in epitaxial growth and removal},
World Scientific, Singapore.

\noindent Ratsch, C. and Scheffler, M. 1998.
Density-functional theory calculations of hopping rates of surface diffusion,
{\em Phys. Rev. B 58, 13163}.

\noindent Reuter, K. and Scheffler, M. 2002.
Composition, structure, and stability of RuO$_{2}$(110) as a function
 of oxygen pressure,
{\em Phys. Rev. B 65, 035406}.

\noindent Reuter, K. and Scheffler, M. 2003a.
First-principles atomistic thermodynamics for oxidation catalysis: Surface phase diagrams and catalytically interesting regions,
{\em Phys. Rev. Lett. 90, 046103}.

\noindent Reuter, K. and Scheffler, M. 2003b.
Composition and structure of the RuO$_2$(110) surface in an O$_2$ and CO environment: Implications for the catalytic formation of CO$_2$,
{\em Phys. Rev. B 68, 045407}.

\noindent Reuter, K. and Scheffler, M. 2004a.
Oxide formation at the surface of late 4$d$ transition metals: Insights from first-principles atomistic thermodynamics,
{\em Appl. Phys. A 78, 793}.

\noindent Reuter, K. and Scheffler, M. 2004b.
Thin Film Nanocatalysis: Oxide Formation at the surface of late transition
Metals. In: Heiz, U., Hakkinen, H. and Landman, U. (eds.) {\em Nanocatalysis: Principles, methods, case Studies}.

\noindent Reuter, K., Frenkel, D. and Scheffler, M. 2004.
The steady state of heterogeneous catalysis, studied with first-principles statistical mechanics,
{\em Phys. Rev. Lett. (submitted)}.

\noindent Ruggerone, P., Ratsch, C. and Scheffler, M. 1997.
Density-functional theory of epitaxial growth of metals. In: King, D.A. and Woodruff, D.P. (eds.) {\em Growth and properties of ultrathin epitaxial layers. The chemical physics of solid surfaces}, vol. 8, 
Elsevier, Amsterdam.

\noindent Sanchez, J.M., Ducastelle, F. and Gratias, D. 1984.
Generalized cluster description of multicomponent systems,
{\em Physica A 128, 334}.

\noindent Scheffler, M. 1988.
Thermodynamic aspects of bulk and surface defects -- first-principles calculations. In: Koukal, J. (ed.) {\em Physics of solid surfaces - 1987}, Elsevier, Amsterdam.

\noindent Scheffler, M. and Dabrowski, J. 1988.
Parameter-free calculations of total energies, interatomic forces, and
vibrational entropies of defects in semiconductors,
{\em Phil. Mag. A 58, 107}.

\noindent Scheffler, M. and Stampfl, C. 2000.
Theory of adsorption on metal substrates. In: Horn, K. and Scheffler, M. (eds.) 
{\em Handbook of surface science}, vol. 2: Electronic structure, 
Elsevier, Amsterdam.

\noindent Stampfl, C. and Scheffler, M. 1995.
Theory of alkali metal adsorption on close-packed metal surfaces,
{\em Surf. Rev. Lett. 2, 317}.

\noindent Stampfl, C., Kreuzer, H.J., Payne, S.H., Pfn\"ur, H., and Scheffler, M. 1999a.
First-principles theory of surface thermodynamics and kinetics,
{\em Phys. Rev. Lett. 83, 2993}.

\noindent Stampfl, C., Kreuzer, H.J., Payne, S.H., and Scheffler, M. 1999b.
Challenges in predictive calculations of processes at surfaces: Surface thermodynamics and catalytic reactions,
{\em Appl. Phys. A 69, 471}.

\noindent Stampfl, C., Ganduglia-Pirovano, M.V., Reuter, K., and Scheffler, M.
2002.
Catalysis and corrosion: The theoretical surface-science context,
{\em Surf. Sci. 500, 368}.

\noindent Stull, D.R. and Prophet, H. 1971.
{\em  JANAF thermochemical tables} (2nd edn.),
U.S. National Bureau of Standards, Washington, D.C.

\noindent Todorova, M., Lundgren, E., Blum, V., Mikkelsen, A., Gray, S.,
Gustafson, J., Borg, M., Rogal, J., Reuter, K., Andersen, J. N., and Scheffler, M. 2003.
The Pd(100)-($\sqrt{5} \times \sqrt{5}$)R27$^{\circ}$-O surface oxide revisited,
{\em Surf. Sci. 541, 101}.

\noindent van de Walle, A. and Ceder, G. 2002.
Automating first-principles phase diagram calculations,
{\em J. Phase Equilibria 23, 348}.

\noindent Vineyard, G.H. 1957.
Frequency factors and isotope effects in solid state rate processes,
{\em J. Phys. Chem. Solids 3, 121}.

\noindent Voter, A.F. 1986.
Classically exact overlayer dynamics: Diffusion of rhodium clusters on Rh(100),
{\em Phys. Rev. B 34, 6819}.

\noindent Voter, A.F., Montalenti, F., and Germann, T.C. 2002.
Extending the time scale in atomistic simulation of materials,
{\em Annu. Rev. Mater. Res. 32, 321}.

\noindent Wang, X.-G., Weiss, W., Shaikhutdinov, Sh.K., Ritter, M., Petersen, M., Wagner, F., Schl\"ogl, R., and Scheffler, M.  1998.
The hematite (alpha--Fe$_{2}$O$_{3}$)(0001) surface: Evidence for domains of distinct chemistry,
{\em Phys. Rev. Lett. 81, 1038}. 

\noindent Wang, X.-G., Chaka, A., and Scheffler, M.  2000.
Effect of the environment on Al$_{2}$O$_{3}$(0001) surface structures,
{\em Phys. Rev. Lett. 84, 3650}.

\noindent Wang, F. and Landau, D.P. 2001. 
Efficient, multiple-range random walk algorithm to calculate the density of states,
{\em Phys. Rev. Lett. 86, 2050}.

\noindent Woodruff, D.P. and Delchar, T.A. 1994.
{\em Modern techniques of surface science} (2nd. edn.), 
Cambridge University Press, Cambridge.

\noindent Xiong, G.M., Schwennicke, C., Pfn\"{u}r, H., and Everts, H.-U. 1997.
Phase diagram and phase transitions of the adsorbate system S/Ru(0001): A Monte Carlo study of a lattice gas model, 
{\em Z. Phys. B 104, 529}.

\noindent Zangwill, A. 1988.
{\em Physics at surfaces}, 
Cambridge University Press, Cambridge.

\noindent Zhdanov, V.P. and Kasemo, B. 1998.
Simulation of oxygen desorption from Pt(111),
{\em Surf. Sci. 415, 403}.

\noindent Zunger, A. 1994, 
First principles statistical mechanics of semiconductor alloys and intermetallic compounds. In: {\em Statics and dynamics of alloy phase
transformations}, Turchi, P.E.A. and Gonis, A. (eds.),
NATO ASI Series, Plenum Press, New York.

\end{document}